\documentclass[11pt,twoside,a4paper]{article}

\usepackage{arxiv}

\usepackage[utf8]{inputenc} 
\usepackage[T1]{fontenc}    
\usepackage{hyperref}       
\usepackage{url}            
\usepackage{booktabs}       
\usepackage{amsfonts}       
\usepackage{nicefrac}       
\usepackage{microtype}      
\usepackage{lipsum}		
\usepackage{graphicx}
\usepackage{natbib}
\usepackage{doi}
\usepackage{caption}
\usepackage{subcaption}

\usepackage{mathtools,upgreek,eucal,mathrsfs,enumitem,amsthm,amsmath}
\usepackage{times,epsfig,makeidx,amsfonts,amsmath,dcolumn,multirow,amssymb,colortbl,bm,url}
\usepackage{algorithm}
\usepackage[]{algorithmic}
 \usepackage[flushleft]{threeparttable}

\newcommand{\btheta}{\bm{\theta}}

\newcommand{\bx}{{\bf x}}

\newcommand{\MOOMIN}{{\text{\tiny MOOMIN}}}
\newcommand{\DIMOM}{{\text{\tiny DIMOM}}}

\newcommand{\D}{{\mathcal D}}

\newcommand{\R}{{\mathbb R}}

\newtheorem{proposition}{Proposition}

\title{An objective non-local prior for skew-symmetric models}

\author{
	\href{https://orcid.org/0000-0001-7183-8407}{\includegraphics[scale=0.06]{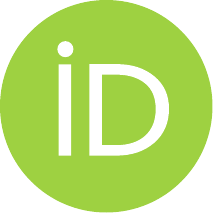}\hspace{1mm} F. Javier Rubio} \\
	Department of Statistical Science\\
	University College London \\
	London, UK\\
	\texttt{f.j.rubio@ucl.ac.uk} 
	}

\renewcommand{\shorttitle}{New Document}

\hypersetup{
pdftitle={XXX},
pdfsubject={stat.ME, stat.AP},
pdfauthor={XXX, Rubio},
}

\begin{document}
\maketitle

\begin{abstract}
We propose an objective non-local prior for testing symmetry against skew-symmetric alternatives. The prior is derived through a formal construction rule by assigning a uniform distribution to a discrepancy-based measure of the shape parameter's effect. This approach avoids the need for user-specified hyperparameters and produces a weakly informative prior tailored to the skew-symmetric family. We illustrate the use of the proposed prior in the context of testing normality against skew-normal alternatives through both a simulation study and a real-data application. 
\end{abstract}


\section{Introduction}\label{sec:intro}

The Bayesian framework constitutes a central paradigm for hypothesis testing and model selection in statistical science. A key challenge within this framework is the specification of a prior distribution, which plays a fundamental role in both tasks.
A substantial body of work has focused on developing so-called \textit{objective priors}, which are priors constructed according to formal principles or rules \citep{kass:1996}. Some examples include Jeffreys’ prior \citep{jeffreys:1946}, reference priors \citep{berger:2009}, and the Wasserstein prior \citep{li:2022}, among many others derived from formal criteria \citep{leisen:2020}. These priors are primarily designed to be weakly informative for estimation and inference on the model parameters, but not necessarily tailored for hypothesis testing.
The impact of prior choice in hypothesis testing and model selection is known to persist even asymptotically. In particular, priors determine the rate of convergence of Bayes factors, which in turn governs the asymptotic behaviour of posterior model probabilities \citep{johnson:2010}. This insight has motivated the development of priors explicitly tailored to control the asymptotic behaviour of Bayes factors, often classified as \textit{local} and \textit{non-local} priors \citep{johnson:2010}. As argued by \cite{johnson:2010}, the use of local priors may be misaligned with the objectives of hypothesis testing, since such priors place substantial mass near the null hypothesis (or baseline model), inducing an imbalance in the ability to detect true alternatives relative to true null hypotheses.
To address this issue, \cite{johnson:2010} introduced the class of non-local priors, which assign zero prior density at the null hypothesis and exclude prior mass in a neighbourhood of the null. This construction penalises the alternative hypothesis when it is nearly indistinguishable from the null, improving the calibration and asymptotic behaviour of Bayes factors under both the null and the alternative.
Notably, objective prior construction strategies, including those that have been specifically proposed for hypothesis testing \citep{berger:1996}, lead to local priors. As a result, existing objective priors do not, in general, address the imbalance induced by local priors in Bayesian testing and model selection.

In this work, we propose a strategy for constructing non-local priors for the shape (skewness) parameter, $\lambda \in \R$, in skew-symmetric models \citep{azzalini:2013}. The prior is defined through an interpretable discrepancy-based measure of the effect of the shape parameter \citep{wagenmakers:2025}, and is identically zero under the null hypothesis of symmetry $\lambda = 0$. By excluding prior mass in a neighbourhood of this null value, the induced prior on the shape parameter is non-local and naturally aligned with the objectives of Bayesian hypothesis testing and model selection \citep{johnson:2010}.

Section \ref{sec:discrepancy} introduces the discrepancy measure, describes the construction of the proposed non-local prior, and presents a tractable approximation. Section \ref{sec:hypothesis} discusses the use of the proposed prior for testing hypotheses of symmetry against skew-symmetric alternatives and characterises the corresponding Bayes factor rates. Section \ref{sec:simulation} presents a simulation study for testing normality against skew-normal alternatives, while Section \ref{sec:application} provides a real-data application. Section \ref{sec:discussion} concludes with a discussion of the insights gained from this work on constructing and specifying non-local priors for shape parameters and potential extensions.

\section{Objective priors based on a discrepancy measure}\label{sec:discrepancy}

\subsection{The discrepancy measure}
We begin by describing the discrepancy measure of interest and its use in the context of skew-symmetric models. Consider the discrepancy measure proposed by \citep{wagenmakers:2025} between two probability density functions $f_1$ and $f_2$ with support on ${\mathcal X}$,
\begin{eqnarray}\label{eq:discrepancy}
d(f_1 \mid\mid f_2) = \int_{\mathcal{X}} \dfrac{f_1(x)}{f_1(x)+f_2(x)} f_1(x)dx.
\end{eqnarray}
This discrepancy measure is symmetric, $d(f_1 \mid\mid f_2) = d(f_2 \mid\mid f_1)$, and it can be interpreted as the expected probability assigned to model $f_1$, relative to model $f_2$, based on a single observation \citep{wagenmakers:2025}. This interpretation has motivated its application in the context of model selection \citep{datta:2025}.

Consider the family of skew-symmetric models with probability density function:
\begin{eqnarray}\label{eq:skewsymmetric}
s_{f,G}(x \mid \mu,\sigma,\lambda) = \frac{2}{\sigma} f\left(\dfrac{x-\mu}{\sigma}\right)G \left(\lambda \omega\left( \dfrac{x-\mu}{\sigma} \right) \right), \quad x\in \R
\end{eqnarray}
where $f$ denotes a continuous symmetric (about $x=0$) probability density function (pdf) with support on ${\mathbb R}$, $G$ is a cumulative distribution function (cdf) with corresponding continuous symmetric density function $g$ also supported on ${\mathbb R}$, and $\omega: \R \to \R$ is an odd diffeomorphism of $\R$ \citep{azzalini:2003}. The parameter $\mu \in \R$ is a location parameter, $\sigma\in \R_+$ is a scale parameter, and $\lambda\in \R$ is a shape (skewness) parameter. The pdf \eqref{eq:skewsymmetric} reduces to the baseline symmetric model $f$ for $\lambda = 0$. 
This formulation encompasses a wide range of distributions of practical interest, including the skew-normal distribution \citep{azzalini:1985,azzalini:2013}. 
Several objective priors have been proposed for the shape parameter $\lambda$ in skew-symmetric models, including reference and Jeffreys' priors \citep{liseo:2006,rubio:2014b}, distance-based priors \citep{dette:2018}, and priors based on the Wasserstein information matrix \citep{li:2022}. All of these constructions lead to local priors, in the sense that they assign positive prior density at the null value $\lambda = 0$.



\subsection{Moment-Objective Minimum-Discrepancy (MOOMIN) Prior}
\begin{figure}[h!]
    \centering
\begin{tabular}{c c c}
\includegraphics[width=0.3\textwidth]{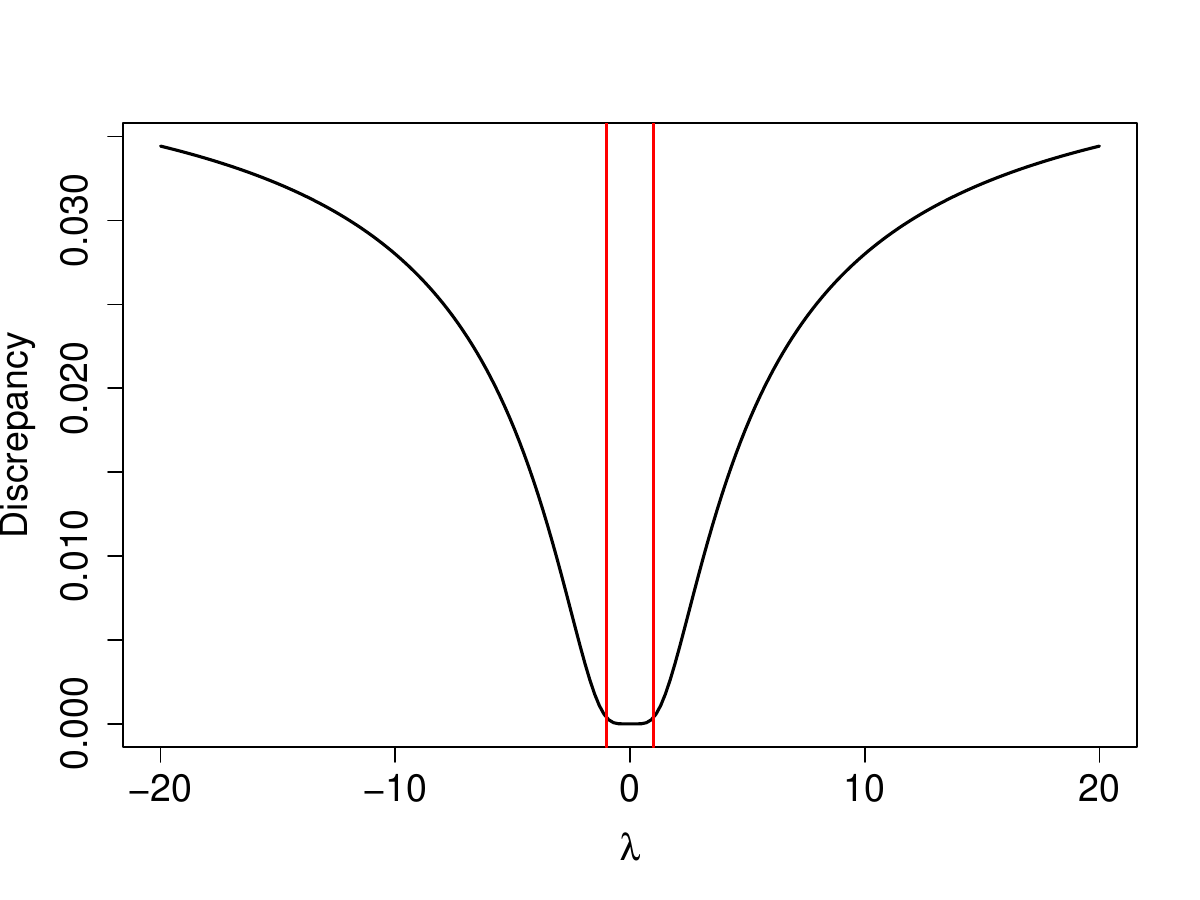} & 
\includegraphics[width=0.3\textwidth]{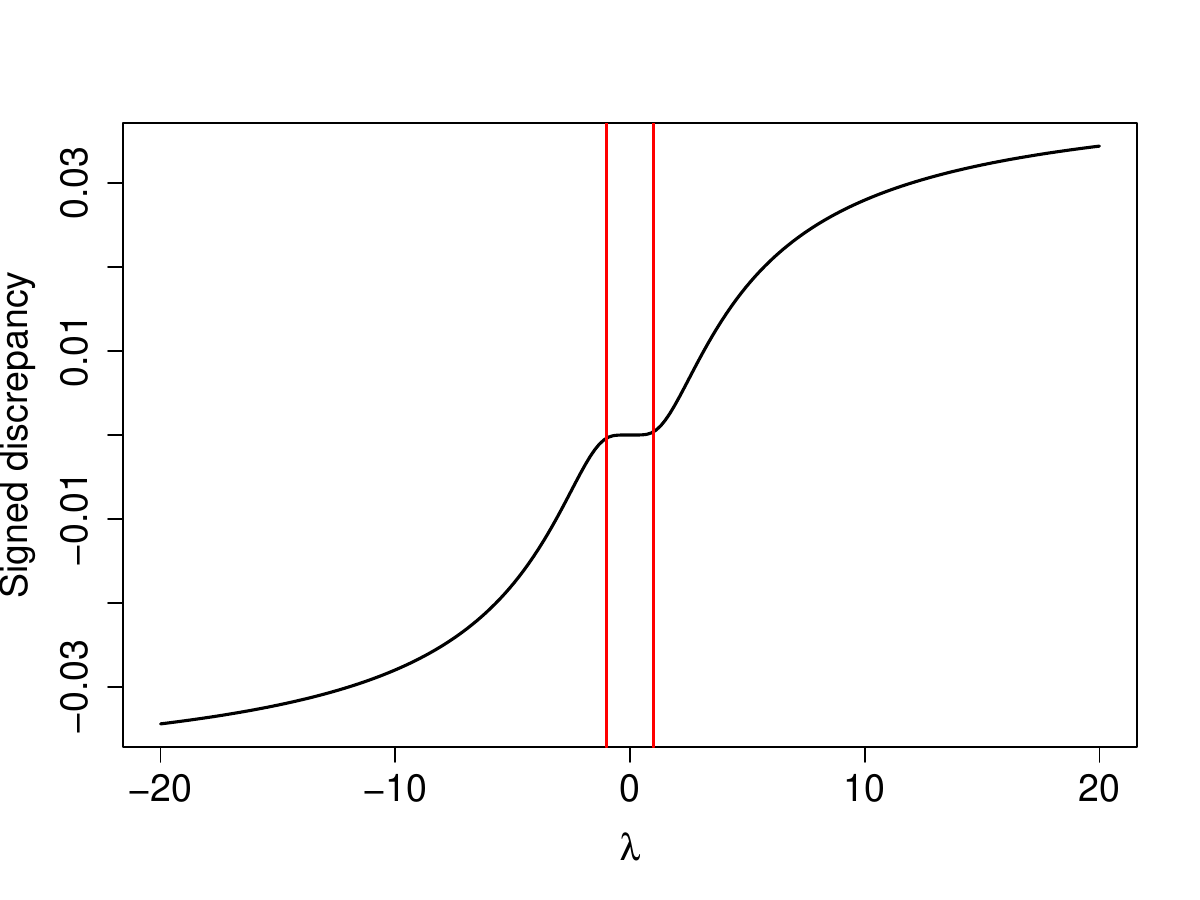} & 
\includegraphics[width=0.3\textwidth]{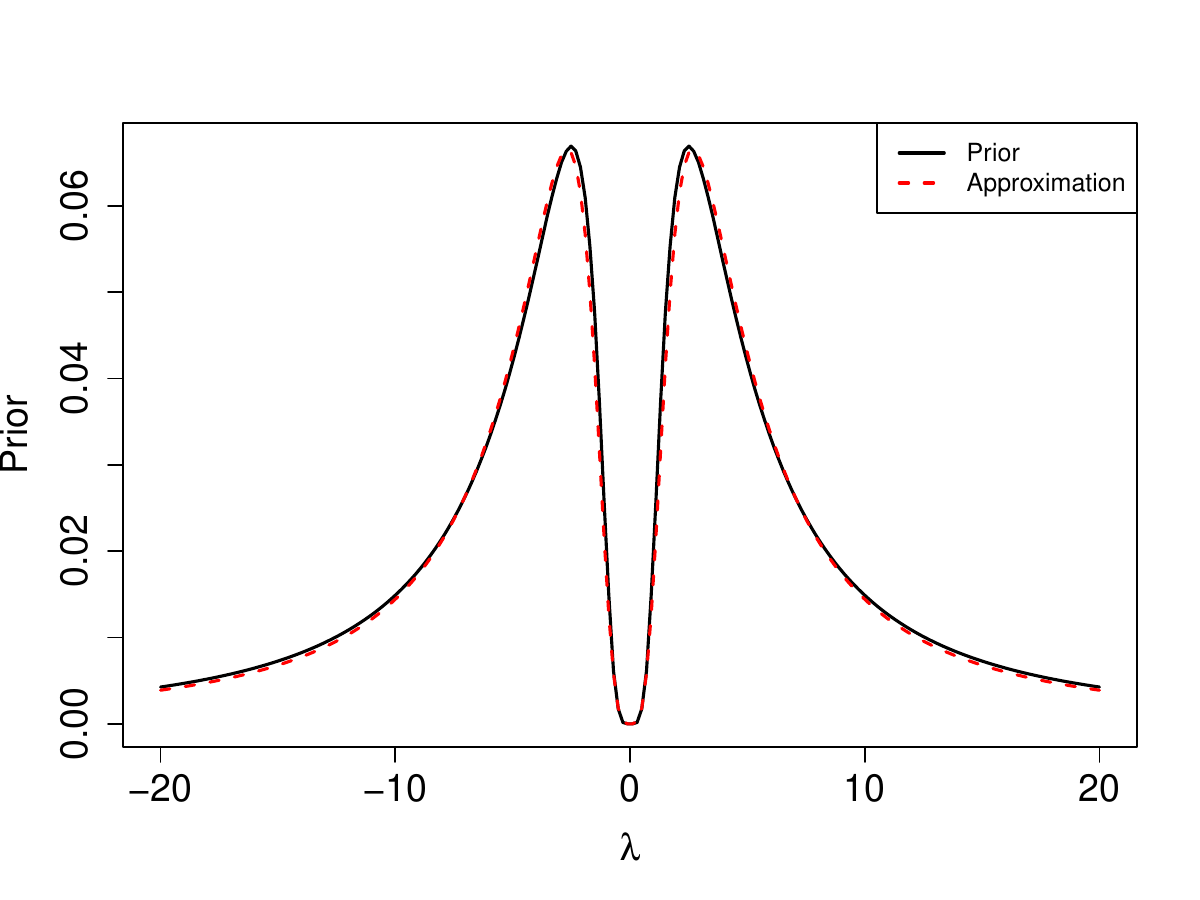} \\
 (a) & (b) & (c)\\
 \end{tabular}
    \caption{Skew-normal distribution: (a) Discrepancy measure, (b) signed discrepancy measure, and (c) MOOMIN prior and its approximation \eqref{eq:moomin_app}. The vertical red lines in (a) and (b) are shown at $\lambda = \pm 1$.}
    \label{fig:sn_min}
\end{figure}

{From standard asymptotic theory \citep{van:2000}, it is known that when a model is misspecified; specifically, if the true data-generating process is the  skew-symmetric density in \eqref{eq:skewsymmetric} but we fit a symmetric location--scale family $f$, the resulting point estimators $(\widehat{\mu}, \widehat{\sigma})$ converge, under mild regularity conditions, to the ``pseudo-true'' values $(\mu^*, \sigma^*)$. These values are defined as the parameters that minimise the Kullback--Leibler divergence (or a related discrepancy) between the symmetric family and the true skew-symmetric model. This asymptotic behaviour implies that the relevant ``competitor'' to the skew-symmetric model is not simply the standard symmetric density $f(x)$, but rather the best-fitting member of the symmetric location--scale family. This motivates the definition of the minimum discrepancy, $\D_{\min}(\lambda)$, which quantifies the irreducible distance between the skew-symmetric model and its closest symmetric approximation:
\begin{eqnarray}\label{eq:discss_min}
\D_{\min}(\lambda) = \inf_{\mu,\sigma} \int_{-\infty}^{\infty}  \frac{\frac{1}{\sigma^2}f\left(\dfrac{x-\mu}{\sigma}\right)^2}{\frac{1}{\sigma}f\left(\frac{x-\mu}{\sigma}\right)+2f(x)G(\lambda \omega(x) )} \, dx,  
\end{eqnarray}
By taking the infimum over $(\mu, \sigma)$, we ensure that the discrepancy reflects the ``smallest'' structural difference between the models rather than a mere difference in location or scale.
Intuitively, this defines $\D_{\min}(\lambda)$ as the discrepancy between the skew-symmetric model and its \textit{projection} onto the symmetric location-scale distributions.}
This discrepancy measure takes values in $[1/2,1/2+C]$, for some constant $0 < C \leq \frac{1}{2}$ dependent on the models being compared. For instance, for the skew-normal distribution, when $f$ and $G$ are the standard normal pdf and cdf, respectively, this discrepancy measure takes values in $[0.5,\,0.5417]$, attaining its minimum at $\lambda = 0$ and its maximum as $\lambda \to \pm \infty$. 
To obtain an injective increasing function of the parameter $\lambda$, we define the signed discrepancy measure:
\begin{equation}\label{eq:signeddisc_min}
M_{\D_{\min}}(\lambda) = \begin{cases}
\frac{1}{2} - \D_{\min}(\lambda) & \textit{ if } \lambda \leq 0\\
\D_{\min}(\lambda) - \frac{1}{2} & \textit{ if } \lambda > 0.
\end{cases}
\end{equation}
which takes values in $[-C,C]$, for some $C>0$ dependent on the choice of $f$ and $G$.
We then propose assigning a uniform prior to $M_{\D_{\min}}(\lambda)$, reflecting weak prior information about the effect of this parameter, leading to the prior density function:
\begin{eqnarray}\label{eq:moomin}
\pi_{\MOOMIN}(\lambda) \propto \frac{d}{d\lambda} M_{\D_{\min}}(\lambda).
 \end{eqnarray}
We refer to prior \eqref{eq:moomin} as the Moment-Objective Minimum-Discrepancy (MOOMIN) prior, due to its connection with non-local MOM priors \citep{johnson:2010}, as we will demonstrate next. The following result characterises some properties of this prior.
\begin{proposition}\label{prop:characterisation_min}
Consider the skew-symmetric model \eqref{eq:skewsymmetric}, and suppose that $\int_{-\infty}^{\infty} \vert \omega(x)\vert f(x) dx < \infty$, $f(0) < \infty$, $\int_{-\infty}^{\infty} \vert \omega(x) \vert g(x) dx < \infty$, and $g$ is continuously differentiable. Then, 
\begin{itemize}
\item[(i)] The prior \eqref{eq:moomin} can be written as:
\begin{eqnarray}\label{eq:priorss_min}
\pi_{\MOOMIN}(\lambda) \propto  \left\vert \int_{-\infty}^{\infty}  \dfrac{\frac{1}{{\sigma^*}^2}f\left(\dfrac{x-\mu^*}{{\sigma^*}}\right)^2 2 \omega(x) f(x)g(\lambda \omega(x) )}{\left [\dfrac{1}{{\sigma^*}}f\left(\frac{x-\mu^*}{{\sigma^*}}\right)+2f(x)G(\lambda \omega(x) )\right]^2} dx \right \vert,
\end{eqnarray}
where $(\mu^*,\sigma^*)$ denote the values obtained from the optimisation in \eqref{eq:discss_min} for each value of $\lambda$.
\item[(ii)]  $\pi_{\MOOMIN}(0) = 0$.
    \item[(iii)] $\pi_{\MOOMIN}(\lambda)$ is symmetric around $\lambda = 0$.
\end{itemize}
Consider now the case $\omega(x) = x$. Then,
\begin{itemize}    
    \item[(iv)] As $\vert \lambda \vert \to \infty$, the tails of the prior $\pi_{\MOOMIN}(\lambda)$ are of order $O(|\lambda|^{-2})$.
\end{itemize}
\end{proposition}
A full characterisation of the vanishing rate of this prior as $\lambda \approx 0$ is challenging, as it depends on the choice of $G$, although numerical methods can be employed to approximate such rates using a polynomial expansion. While the prior does not admit a closed-form expression, it remains numerically tractable, as it only requires one-dimensional integration. To facilitate a faster and more practical implementation, we consider an empirical non-local moment Student-$t$ approximation of the form:
\begin{equation}\label{eq:moomin_app}
\pi_{\MOOMIN}^{\text{app}}(\lambda) =  \dfrac{\vert \lambda \vert^k}{M(1 + a\lambda^2)^m},
\end{equation}
where $M$ is the normalising constant, $a,k>0$, and $m > \frac{k+1}{2}$. As $\vert \lambda \vert \to \infty$, the tails of the approximate prior $\pi_{\D_{\min}}(\lambda)$ are of order $O(|\lambda|^{k-2m})$.
As $\vert \lambda \vert \to 0$, the approximate prior $\pi_{\MOOMIN}^{\text{app}}(\lambda)$ vanishes at a rate $O(|\lambda|^k)$.
For the skew-normal distribution case, the choice $k=4$,  $m = 3$, $a=0.28$ leads to the approximation presented in Figure \ref{fig:sn_min}. The value $k=4$ is obtained by finding the least squares approximation of the polynomial $\vert \lambda \vert^4$ to the MOOMIN prior \eqref{eq:moomin} in a neighbourhood of $\lambda = 0$.
{The Appendix shows two more examples using the skew-logistic and skew-$\operatorname{sech}$ distributions, highlighting the different interpretations and roles of the parameter $\lambda$ in different members of the skew-symmetric family of distributions.}

\subsection{Discrepancy-Informed Moment (DIMOM) Prior}
Having established the connection between the proposed discrepancy-based priors and non-local priors in the preceding sections, we now consider the construction of a moment (MOM) prior \citep{johnson:2010}, calibrated using the information provided by the discrepancy measure in \eqref{eq:discss_min}. From Figure \ref{fig:sn_min}, we observe that the region $\vert \lambda \vert \leq 1$ has a negligible effect on the skewness of the distribution. As an empirical guideline, we define a moment-normal prior that assigns low probability mass to this region, leading to the following heuristic choice:
\begin{equation}\label{eq:DIMOM}
\pi_{\DIMOM}(\lambda) = \dfrac{\lambda^2}{\sqrt{2\pi}\sigma_M}\exp\left( - \dfrac{\lambda^2}{2\sigma_M^2} \right).
\end{equation}
We consider the choice $\sigma_M = 1.69$, which results in a prior mass of approximately $5\%$ on the region $\vert \lambda \vert \leq 1$. We refer to this prior as the discrepancy-informed moment prior (DIMOM). The idea of calibrating MOM priors for shape parameters based on their effect on the baseline distribution has been applied to other distributions in \cite{rossell:2018}. Although \eqref{eq:DIMOM} defines a non-local prior, it imposes a milder reduction of prior mass around the null hypothesis, and its tails are lighter than those of the MOOMIN prior in \eqref{eq:moomin}.

\section{Hypothesis testing}\label{sec:hypothesis}

Based on the results in Section \ref{sec:discrepancy}, the prior distribution in \eqref{eq:moomin} can be interpreted as a non-local prior \citep{johnson:2010} for testing hypotheses of the form:
\begin{equation}\label{eq:hypothesis}
 H_0: \lambda = 0 \quad \text{vs.} \quad H_1 : \lambda \neq 0.
\end{equation}
Consequently, one could view these priors as objective priors (in the sense that they are obtained through a formal rule) for testing symmetry against skew-symmetric alternatives. A natural question is thus what are the asymptotic properties of the Bayes factors obtained with these priors, and how to operationalise their use in practice. The following subsections address these questions using the Laplace and integrated Laplace approximations to the marginal likelihood.
\subsection{Bayes factor rates}
Let $\bx = \{x_1,\dots,x_n\}$ be independent and identically distributed (\textit{i.i.d.}) samples from the skew-symmetric model in \eqref{eq:skewsymmetric}. Denote by $\ell_1(\mu,\sigma,\lambda)$ the log-likelihood function corresponding to the skew-symmetric model \eqref{eq:skewsymmetric}, and by $\ell_0(\mu,\sigma)$ the log-likelihood function corresponding to the baseline symmetric model $f$. Let $BF(1 \mid 0)$ denote the Bayes factor comparing the skew-symmetric model \eqref{eq:skewsymmetric} with the baseline symmetric model $f$, that is,

\begin{equation*}
BF(1 \mid 0) = \dfrac{p(\bx \mid 1)}{p(\bx \mid 0)} = \dfrac{\int \exp\{\ell_1(\mu,\sigma,\lambda)\} \pi_1(\mu,\sigma,\lambda) \, d\mu d\sigma d\lambda}{\int \exp\{\ell_0(\mu,\sigma)\} \pi_0(\mu,\sigma) \, d\mu d\sigma},
\end{equation*}
where $\pi_1(\mu,\sigma,\lambda)$ and $\pi_0(\mu,\sigma)$ are the corresponding priors, $p(\bx \mid 1)$ and $p(\bx \mid 0)$ denote the marginal likelihoods of the data for the skew-symmetric and baseline symmetric models, respectively. The posterior probability of model ${\mathcal M} = 1,0$ can be obtained as
\begin{equation*}
\Pr({\mathcal M} = j \mid \bx ) = \dfrac{p(\bx \mid {\mathcal M} = j) \pi({\mathcal M} = j)}{p(\bx \mid {\mathcal M} = 1) \pi({\mathcal M} = 1) + p(\bx \mid {\mathcal M} =  0)\pi({\mathcal M} = 0)}, \quad j = 1,0,
\end{equation*}
where $\pi({\mathcal M} = j)$ denotes the prior probability of model $j$.
For a general choice of prior, the marginal likelihoods $p(\bx \mid {\mathcal M} = j)$ are not available in closed form, and thus require numerical integration. As a practical strategy for approximating this quantity, we consider a Laplace approximation.
\begin{equation*}
\widehat{p}(\bx \mid \mathcal{M})= \exp\{\ell_{\mathcal{M}}(\tilde{\btheta}_{\mathcal{M}}) + \log \pi_{\mathcal{M}}(\tilde{\btheta}_{\mathcal{M}}) \}
(2\pi)^{d_{\mathcal{M}}/2} \det \left( H_{\mathcal{M}}(\tilde{\btheta}_{\mathcal{M}}) + \nabla^2 \log\pi_{\mathcal{M}}(\tilde{\btheta}_{\mathcal{M}}) \right)^{-1/2},
\end{equation*}
where $\btheta_{\mathcal{M}}$ are the parameters associated with model $\mathcal{M}$, $\tilde{\btheta}_{\mathcal{M}}$ denotes the maximum a posteriori (MAP) estimate under the corresponding prior,
$H_{\mathcal{M}}(\tilde{\btheta}_{\mathcal{M}})$ is the Hessian matrix of the log-likelihood function evaluated at the MAP, and
$\nabla^2 \log\pi_{\mathcal{M}}(\tilde{\btheta}_{\mathcal{M}})$ is the Hessian matrix of the log-prior evaluated at the MAP. Let us denote the Bayes factor, based on the Laplace approximation, as $\widehat{BF}(1 \mid 0)$. The following result provides the convergence rates of the approximate Bayes factor $\widehat{BF}(1 \mid 0)$.
\begin{proposition}
Let $\bx = \{x_1,\dots,x_n\}$ be an \textit{i.i.d.} sample from a skew-symmetric model of the form \eqref{eq:skewsymmetric}. Suppose that the functions $f$ and $G$ satisfy the conditions stated in Proposition \ref{prop:characterisation_min}. Consider testing a hypothesis of the form \eqref{eq:hypothesis} using the prior $\pi(\mu,\sigma) \propto \dfrac{1}{\sigma}$ under $H_0$, and the prior $\pi(\mu,\sigma,\lambda) \propto \dfrac{1}{\sigma}\,\pi_{\lambda}(\lambda)$ under $H_1$. Then the following results hold.

\begin{itemize}
    \item[(i)] If the null hypothesis is true, the approximate Bayes factors, under Jeffreys, MOOMIN, and DIMOM priors, satisfy
    \begin{equation*}
        \widehat{BF}(1 \mid 0) = O_p(n^{-k}),
    \end{equation*}
    where $k = 1/2$ for local priors $\pi_{\lambda}(\lambda)$. For non-local priors $\pi_{\lambda}(\lambda)$, $k = 3/2$ for the DIMOM prior \eqref{eq:DIMOM}, and $k = 5/2$ for the approximate MOOMIN prior \eqref{eq:moomin_app}.

    \item[(ii)] If the alternative hypothesis is true, the approximate Bayes factor satisfies
    \begin{equation*}
        \frac{1}{n}\log \left\{\widehat{BF}(1 \mid 0)\right\} \stackrel{\Pr}{\to} c_{\btheta_0},
    \end{equation*}
    where $c_{\btheta_0}$ is a constant that depends on the true value of the parameter $\btheta_0 = (\mu_0,\sigma_0,\lambda_0)$.
\end{itemize}
\end{proposition}
This result illustrates the accelerated convergence of the Bayes factor in detecting true null hypotheses under non-local priors, which is induced by assigning zero prior density and reducing prior mass in a neighbourhood of $\lambda=0$. This effectively penalises small deviations from the null more severely than standard local priors, leading to a faster growth of the Bayes factor in favour of $H_0$. The MOOMIN prior (and its approximation) is a non-local prior \eqref{eq:DIMOM} for the parameter $\lambda$ relative to the baseline symmetric family, removing prior mass near $\lambda = 0$ at a faster rate. 

\section{Simulation Study}\label{sec:simulation}
\begin{figure}[h!]
    \centering
\begin{tabular}{c c c}
\includegraphics[width=0.3\textwidth]{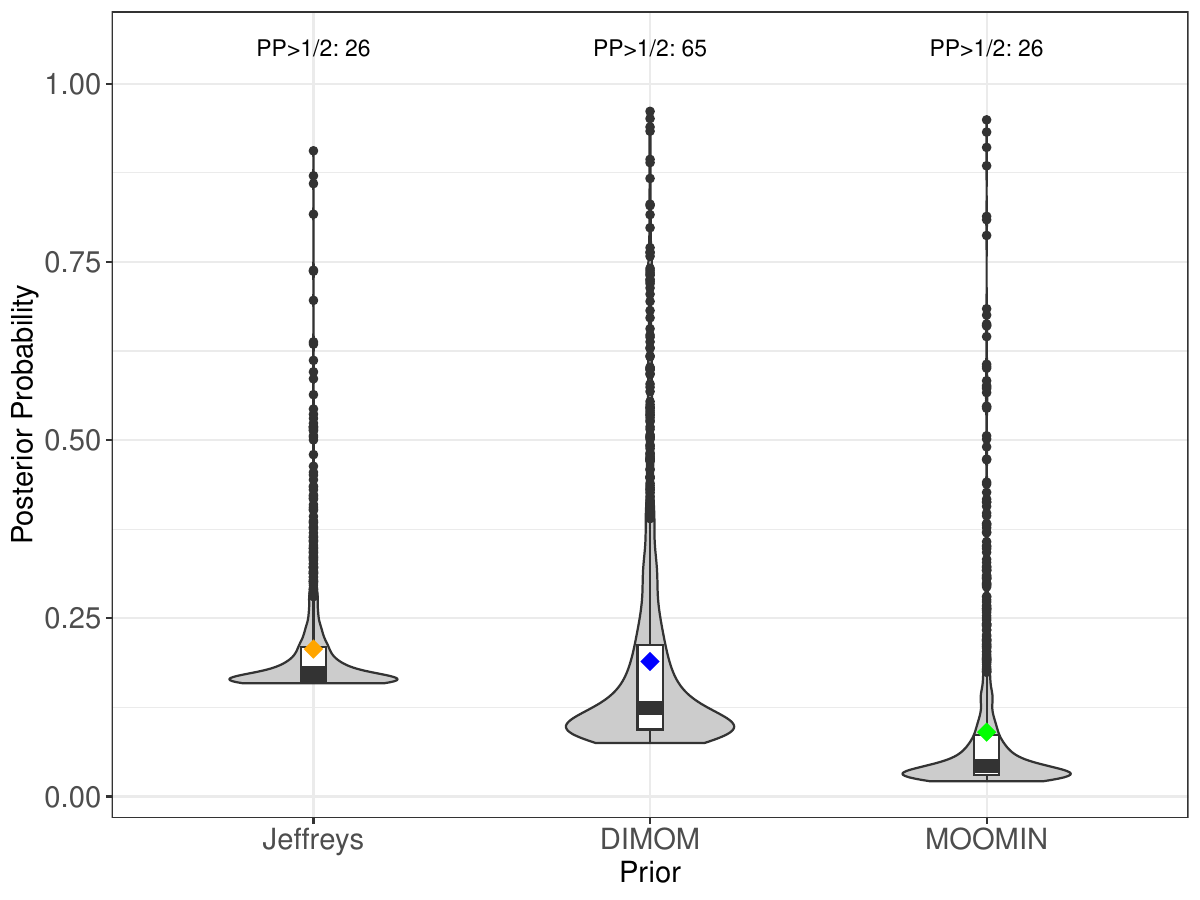} & 
\includegraphics[width=0.3\textwidth]{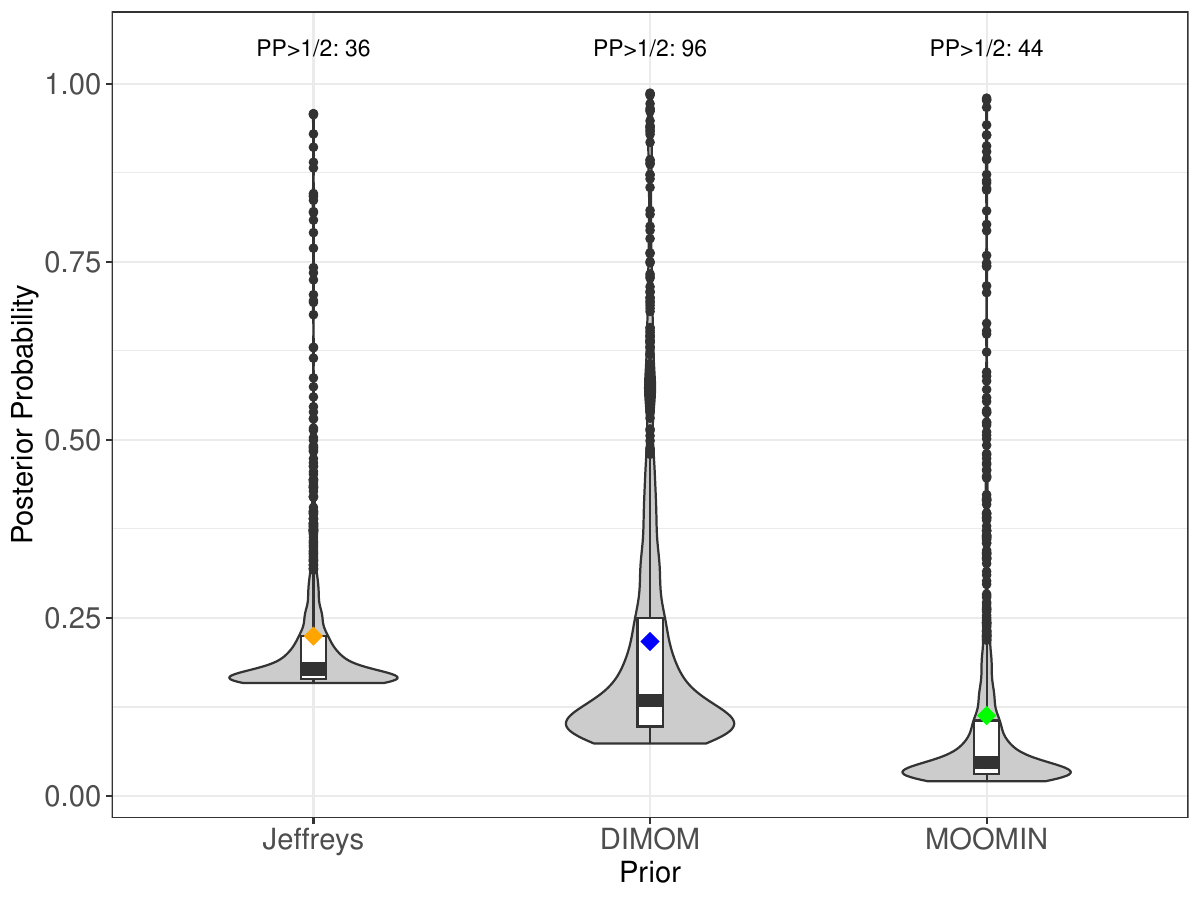} & 
\includegraphics[width=0.3\textwidth]{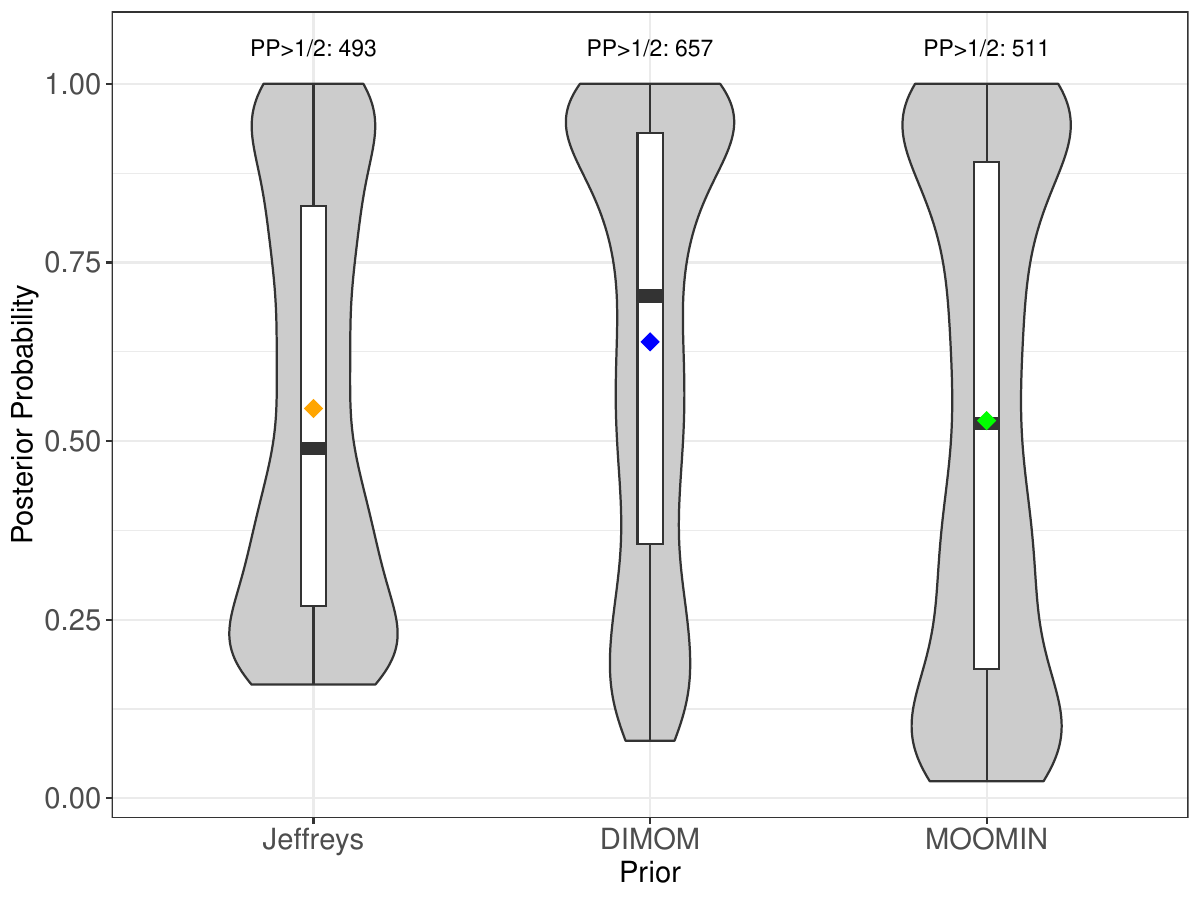} \\
 (a) & (b) & (c)\\
 \end{tabular}
    \caption{Simulation results for $n=100$ based on the Jeffreys prior, the approximate MOOMIN prior, and the DIMOM prior: (a) $\lambda = 0$, (b) $\lambda = 1$, and (c) $\lambda = 2.5$.}
    \label{fig:sn_sim_n100}
\end{figure}

We now present a simulation study to assess the performance of the proposed priors in testing normality against skew-normal alternatives. Specifically, we assume that $f$ and $G$ are the standard normal pdf and cdf, respectively, and consider testing \eqref{eq:hypothesis}. For the skew-normal model, we adopt the prior structure 
$\pi_1(\mu,\sigma,\lambda) = \frac{1}{\sigma} \pi_{\lambda}(\lambda)$, while for the normal null model we use the reference prior $\pi_0(\mu,\sigma) = \frac{1}{\sigma}$.
For the prior $\pi_{\lambda}(\lambda)$, we consider three choices: (i) the Jeffreys prior approximated with the Student-$t$ distribution with $1/2$ degrees of freedom \citep{rubio:2014b}, (ii) the approximate MOOMIN prior \eqref{eq:moomin_app}, and (iii) the DIMOM prior \eqref{eq:DIMOM}. We simulate $N = 1{,}000$ samples of sizes $n = 50, 100, 200,$ and $500$ from the skew-normal distribution with $\lambda = 0$ (normality), $\lambda = 1$ (close to normality), and $\lambda = 2.5$ (a clear departure from normality). For each sample, we compute the posterior probability of the skew-normal model using an integrated (nested) Laplace approximation (ILA), where the parameters $(\mu,\sigma)$ are integrated using a Laplace approximation and the parameter $\lambda$ is integrated using quadrature methods. The marginal likelihood for the normal model is obtained in closed-form. This method does not rely on the posterior of $\lambda$ being log-concave and typically leads to more accurate approximations. We assume equal prior probabilities for both models.
Results for $n = 100$ are presented in Figure \ref{fig:sn_sim_n100}, showing that the proposed priors (approximate MOOMIN and DIMOM) assign lower posterior probabilities to the skew-normal model when the null hypothesis is true or when $\lambda = 1$, as this model is very close to normality. In contrast, when the true model exhibits a clear departure from normality ($\lambda = 2.5$), the proposed non-local priors remain competitive with local priors despite imposing a stronger penalty on the alternative. These patterns become even more pronounced in larger samples, as shown in the Appendix.

\section{Real data application}\label{sec:application}
We now illustrate the use of the proposed priors for testing normality using the Australian athletes data set (\texttt{sn} R package). We analyse the Body Mass Index (BMI $=$ weight (kg) / height (m)$^2$) of $n = 100$ Australian female athletes using the Jeffreys prior approximated with the Student-$t$ distribution with $1/2$ degrees of freedom \citep{rubio:2014b}, the approximate MOOMIN prior \eqref{eq:moomin_app}, and the DIMOM prior \eqref{eq:DIMOM}. The BIC for the skew-normal model is $484.82$, while the BIC for the normal model is $486.15$, indicating that the data slightly favour the skew-normal model. The Shapiro-Wilk test yields a p-value of $0.034$, rejecting normality at the $5\%$ significance level, although it remains relatively close to the nominal threshold. The posterior probabilities for the skew-normal model are shown in Table \ref{tab:aispp}, with all priors assigning probabilities greater than $0.5$, and the DIMOM prior leading to the largest probability. This illustrates that normal tails and the squared-rate vanishing property of the DIMOM prior may not be sufficient to enforce parsimony in this context.  

This data set is known to include an outlier, corresponding to an individual with BMI greater than $30$. We therefore identify outliers using a robust detection method \citep{leys:2013} based on the absolute deviation around the median. This procedure identifies a single outlier, which is then removed, and the normality tests are repeated. After removal, the BIC for the skew-normal model is $469.62$, while the BIC for the normal model is $466.89$, indicating that the data now slightly favour the normal model. The Shapiro–Wilk test yields a p-value of $0.607$, which does not reject normality at the $5\%$ significance level and is far from the nominal threshold. The posterior probabilities for the skew-normal model are shown in Table \ref{tab:aispp}, with all priors assigning probabilities well below $0.5$, and the MOOMIN prior producing the lowest posterior probability.

\begin{table}[ht]
\centering
\begin{tabular}{cccc}
  \hline
 & Jeffreys & DIMOM & MOOMIN  \\ 
  \hline
Posterior Probability ($H_1$) (full data) & 0.52 & 0.73 & 0.52  \\ 
Posterior Probability ($H_1$) (no outlier) & 0.23 & 0.25 & 0.11 \\  
   \hline
\end{tabular}
\caption{Australian athletes data: Posterior probabilities of the skew-normal model.}
\label{tab:aispp}
\end{table}

\section{Discussion}\label{sec:discussion}
We have proposed a formal rule for constructing objective priors for testing symmetry against skew-symmetric alternatives, and illustrated their use in testing normality against skew-normal alternatives. Beyond the contribution in terms of constructing and characterising the proposed objective non-local priors, this work provides additional insights into the design of non-local priors for shape parameters. First, it is crucial to account for the effect of the shape parameter relative to the null model in order to define the vanishing rate and the allocation of prior mass to regions around the null model where the shape parameter has a negligible effect. When specifying a weakly informative prior, it is necessary to consider heavier-than-normal tails in the context of skew-symmetric models. This observation aligns with \cite{wu:2020}, who argue that scale mixtures of non-local priors can be beneficial when little prior information is available about the effect size.

{The proposed prior construction can be directly extended to other flexible distributions with a skewness parameter \citep{jones:2015}. For instance, the Appendix shows that this extension can be done effectively for two-piece distributions \citep{rubio:2020}.
Extending the proposed prior construction to other types of distributions (\textit{e.g.}~with positive or bounded support) is feasible, provided one can define an appropriate nested sub-model and a suitable one-to-one discrepancy measure.
Another possible extension of the proposed prior construction involves assigning a $\mathrm{Beta}(\alpha,\beta)$ prior to the (suitably transformed) signed discrepancy measure in \eqref{eq:signeddisc_min}. This modification changes the tail behaviour of the resulting prior, but its characterisation remains straightforward using the results in \cite{dette:2018}. 
Extensions to models with two or more shape parameters would also be of interest, but remain a more substantial challenge.}

R codes and data are available at \url{https://github.com/FJRubio67/MOOMIN}.

\bibliographystyle{plainnat}
\bibliography{references} 

@article{rubio:2020,
  title={The family of two-piece distributions},
  author={Rubio, F.J. and Steel, M.F.J.},
  journal={Significance},
  volume={17},
  number={1},
  pages={12--13},
  year={2020}
}

@article{wu:2020,
  title={Hyper nonlocal priors for variable selection in generalized linear models},
  author={Wu, H.H. and Ferreira, M.A.R. and Elkhouly, M. and Ji, T.},
  journal={Sankhya A},
  volume={82},
  number={1},
  pages={147--185},
  year={2020}
}

@article{berger:2009,
  title={The formal definition of reference priors},
  author={Berger, J.O. and Bernardo, J.M. and Sun, D.},
  journal={The Annals of Statistics},
  volume={37},
  number={2},
  pages={905--938},
  year={2009}
}

@book{van:2000,
  title={Asymptotic Statistics},
  author={Van der Vaart, A.W.},
  year={2000},
  publisher={Cambridge University Press}
}

@article{datta:2025,
  title={{On Bayes factor functions}},
  author={Datta, S. and Guha, R. and Shudde, R. and Johnson, V.E.},
  journal={Bayesian Analysis},
  year={2025},
  volume={20},
  pages = {1399--1427}
}

@article{wang:2004,
  title={A skew-symmetric representation of multivariate distributions},
  author={Wang, J. and Boyer, J. and Genton, M.G.},
  journal={Statistica Sinica},
  pages={1259--1270},
  year={2004}
}

@article{johnson:2010,
  title={On the use of non-local prior densities in {B}ayesian hypothesis tests},
  author={Johnson, V.E. and Rossell, D.},
  journal={Journal of the Royal Statistical Society: Series B (Statistical Methodology)},
  volume={72},
  number={2},
  pages={143--170},
  year={2010}
}

@article{jeffreys:1946,
  title={An invariant form for the prior probability in estimation problems},
  author={Jeffreys, H.},
  journal={Proceedings of the Royal Society of London. Series A. Mathematical and Physical Sciences},
  volume={186},
  number={1007},
  pages={453--461},
  year={1946}
}

@article{liseo:2006,
  title={A note on reference priors for the scalar skew-normal distribution},
  author={Liseo, B. and Loperfido, N.},
 journal={Journal of Statistical planning and inference},
  volume={136},
  number={2},
  pages={373--389},
  year={2006}
}

@article{rossell:2018,
  title={Tractable {B}ayesian variable selection: beyond normality},
  author={Rossell, D. and Rubio, F.J.},
  journal={Journal of the American Statistical Association},
  volume={113},
  number={524},
  pages={1742--1758},
  year={2018},
  publisher={Taylor \& Francis}
}

@article{berger:1996,
  title={{The intrinsic Bayes factor for model selection and prediction}},
  author={Berger, J.O. and Pericchi, L.R.},
  journal={Journal of the American Statistical Association},
  volume={91},
  number={433},
  pages={109--122},
  year={1996}
}

@article{kass:1995,
  title={Bayes factors},
  author={Kass, R.E. and Raftery, A.E.},
  journal={Journal of the American Statistical Association},
  volume={90},
  number={430},
  pages={773--795},
  year={1995}
}

@article{leys:2013,
  title={Detecting outliers: Do not use standard deviation around the mean, use absolute deviation around the median},
  author={Leys, C. and Ley, C. and Klein, O. and Bernard, P. and Licata, L.},
  journal={Journal of Experimental Social Psychology},
  volume={49},
  number={4},
  pages={764--766},
  year={2013}
}

@misc{wagenmakers:2025,
  title={A Discrepancy Measure Based on Expected Posterior Probability},
  author={Wagenmakers, E.J. and Grasman, R.P.P.P.},
  year = {2025},
  note={Preprint}
}

@article{azzalini:1985,
  title={A class of distributions which includes the normal ones},
  author={Azzalini, A.},
  journal={Scandinavian Journal of Statistics},
  volume={12},
  pages={171--178},
  year={1985}
}

@book{carter:2001,
  title={Foundations of Mathematical Economics},
  author={Carter, M.},
  year={2001},
  publisher={MIT Press}
}

@article{azzalini:2003,
  title={Distributions generated by perturbation of symmetry with emphasis on a multivariate skew-t distribution},
  author={Azzalini, A. and Capitanio, A.},
  journal={Journal of Royal Statistical Society, Series B},
  volume={65},
  pages={367--389},
  year={2003}
}

@book{azzalini:2013,
  title={The skew-normal and related families},
  author={Azzalini, A.},
  volume={3},
  year={2013},
  publisher={Cambridge University Press}
}

@article{kass:1996,
  title={The selection of prior distributions by formal rules},
  author={Kass, R.E. and Wasserman, L.},
  journal={Journal of the American Statistical Association},
  volume={91},
  number={435},
  pages={1343--1370},
  year={1996}
}

@article{rubio:2014,
  title={Inference in two-piece location-scale models with {J}effreys priors (with discussion)},
  author={Rubio, F.J. and Steel, M.F.J.},
  journal={Bayesian Analysis},
  volume={9},
  number={1},
  pages={1--22},
  year={2014}
}

@article{rubio:2014b,
  title={On the independence {J}effreys prior for skew-symmetric models},
  author={Rubio, F.J. and Liseo, B.},
  journal={Statistics \& Probability Letters},
  volume={85},
  pages={91--97},
  year={2014}
}

@article{leisen:2020,
  title={On a Class of Objective Priors from Scoring Rules (with Discussion)},
  author={Leisen, F. and Villa, C. and Walker, S.G.},
  journal={Bayesian Analysis},
  volume={15},
  number={4},
  pages={1345--1423},
  year={2020}
}

@article{jones:2015,
  title={On families of distributions with shape parameters},
  author={Jones, M.C.},
  journal={International Statistical Review},
  volume={83},
  number={2},
  pages={175--192},
  year={2015}
}

@article{dette:2018,
  title={Natural (Non-) Informative Priors for Skew-symmetric Distributions},
  author={Dette, H. and Ley, C. and Rubio, F.J.},
  journal={Scandinavian Journal of Statistics},
  volume={45},
  number={2},
  pages={405--420},
  year={2018}
}

@article{li:2022,
  title={On a prior based on the {W}asserstein information matrix},
  journal={Statistics \& Probability Letters},
  author={Li, W. and Rubio, F.J.},
  volume={190},
  pages={109645},
  year={2022},
  publisher={Elsevier}
}

\clearpage

\section*{Appendix}\label{sec:appendix}

\subsection*{Proof of Proposition 1}
\begin{itemize}
    \item[(i)] Define the function
\begin{eqnarray*}
h(x;\mu,\sigma,\lambda) = \dfrac{f\left(\dfrac{x-\mu}{\sigma}\right)^2}{\sigma f\left(\dfrac{x-\mu}{\sigma}\right)+2\sigma^2f(x)G(\lambda \omega(x) )}
\end{eqnarray*}
Then, we can write the prior as:
\begin{eqnarray*}
\pi_{\D_{\min}}(\lambda) &\propto& \frac{d}{d\lambda} M_{\D_{\min}}(\lambda)  = \left \vert  \frac{d}{d\lambda} \D_{\min}(\lambda) \right \vert \\
&=& \left \vert \frac{d}{d\lambda} \inf_{\mu,\sigma} \int_{-\infty}^\infty h(x;\mu,\sigma,\lambda) \, dx \right \vert
\end{eqnarray*}
Applying the Envelop Theorem \citep{carter:2001} we get that
\begin{eqnarray*}
\pi_{\D_{\min}}(\lambda) &\propto& \frac{d}{d\lambda} M_{\D_{\min}}(\lambda)  = \left \vert  \frac{d}{d\lambda} \D_{\min}(\lambda) \right \vert \\
&=& \left \vert  \int_{-\infty}^\infty \frac{d}{d\lambda} h(x;\mu,\sigma,\lambda) \Bigg \vert_{\mu = \mu^*, \sigma=\sigma^*}\, dx \right \vert\\
 &=&  \left\vert \int_{-\infty}^{\infty}  \dfrac{\dfrac{1}{{\sigma^*}^2}f\left(\dfrac{x-\mu^*}{{\sigma^*}}\right)^2 2 \omega(x) f(x)g(\lambda \omega(x) )}{\left [\dfrac{1}{{\sigma^*}}f\left(\dfrac{x-\mu^*}{{\sigma^*}}\right)+2f(x)G(\lambda \omega(x) )\right]^2} dx \right \vert.  
\end{eqnarray*}

\item[(ii)] Note also that $\lim_{\lambda\to 0} \sigma^*(\lambda) = 1$ and  $\lim_{\lambda\to 0} \mu^*(\lambda) = 0$ as the skew-symmetric pdf $s_{f,G}$ converge to the symmetric baseline pdf $f$ as $\lambda\to0$ \citep{wang:2004}. Thus evaluating the previous equation at $\lambda = 0$, and noting that $\omega(x)$ is an odd function, we obtain
\begin{eqnarray*}
\pi_{\D_{\min}}(0) &\propto&   \left\vert g(0) \int_{-\infty}^{\infty}  \omega(x) f(x) dx \right \vert = 0.  
\end{eqnarray*}

\item[(iii)] Denote the integrand:
\[
I(x,\lambda)=
\frac{
\frac{1}{\sigma^*(\lambda)^2}
f\!\left(\frac{x-\mu^*(\lambda)}{\sigma^*(\lambda)}\right)^2
\,2\omega(x) f(x)\,
g\!\left(\lambda \omega(x)\right)
}{
\left[
\frac{1}{\sigma^*(\lambda)}
f\!\left(\frac{x-\mu^*(\lambda)}{\sigma^*(\lambda)}\right)
+2 f(x) G\!\left(\lambda \omega(x)\right)
\right]^2
}.
\]

Consider the integrand evaluated at $(-x,-\lambda)$.
Using the property $s_{f,G}(x;\mu,\sigma,-\lambda) = s_{f,G}(-x;\mu,\sigma,\lambda)$ for skew-symmetric distributions \citep{azzalini:1985,wang:2004}, it follows that on $\mu^*(-\lambda) = -\mu^*(\lambda)$ and $\sigma^*(-\lambda)=\sigma^*(\lambda)$, and
\[
\frac{-x-\mu^*(-\lambda)}{\sigma^*(-\lambda)}
=
\frac{-x+\mu^*(\lambda)}{\sigma^*(\lambda)}
=
-\frac{x-\mu^*(\lambda)}{\sigma^*(\lambda)}.
\]
Since $f$ is symmetric about $0$, it follows that
\[
f\!\left(\frac{-x-\mu^*(-\lambda)}{\sigma^*(-\lambda)}\right)
=
f\!\left(\frac{x-\mu^*(\lambda)}{\sigma^*(\lambda)}\right).
\]

Moreover, since $\omega$ is odd and $f$ is symmetric,
\[
\omega(-x)f(-x)=-\omega(x)f(x).
\]
Because $g$ is symmetric, $g(-u)=g(u)$, and because $G$ is symmetric,
$G(-u)=1-G(u)$ for all $u\in\mathbb{R}$.
Hence,
\[
g\!\left(-\lambda\omega(-x)\right)
=
g\!\left(\lambda\omega(x)\right),
\qquad
G\!\left(-\lambda\omega(-x)\right)
=
1-G\!\left(\lambda\omega(x)\right).
\]

Combining these identities yields
\[
I(-x,-\lambda)
=
-
\frac{
\frac{1}{\sigma^*(\lambda)^2}
f\!\left(\frac{x-\mu^*(\lambda)}{\sigma^*(\lambda)}\right)^2
\,2\omega(x) f(x)\,
g\!\left(\lambda \omega(x)\right)
}{
\left[
\frac{1}{\sigma^*(\lambda)}
f\!\left(\frac{x-\mu^*(\lambda)}{\sigma^*(\lambda)}\right)
+2 f(x) G\!\left(\lambda \omega(x)\right)
\right]^2
}
=
- I(x,\lambda).
\]

Applying the change of variables $x\mapsto -x$,
\[
\int_{-\infty}^{\infty} I(x,-\lambda)\,dx
=
\int_{-\infty}^{\infty} I(-x,-\lambda)\,dx
=
- \int_{-\infty}^{\infty} I(x,\lambda)\,dx.
\]

Since $\pi_{D_{\min}}(\lambda)$ is defined as the absolute value of this integral,
\[
\pi_{D_{\min}}(-\lambda)
\propto
\left|
- \int_{-\infty}^{\infty} I(x,\lambda)\,dx
\right|
=
\left|
\int_{-\infty}^{\infty} I(x,\lambda)\,dx
\right|
\propto
\pi_{D_{\min}}(\lambda).
\]
This establishes the symmetry of the prior about zero.

\item[(iv)] Let $K$ be the normalising constant of $\pi_{\MOOMIN}(\lambda)$. Then, we can see that

\begin{eqnarray*}
\pi_{\MOOMIN}(\lambda) &=&  K\left\vert \int_{-\infty}^{\infty}  \dfrac{\dfrac{1}{{\sigma^*}^2}f\left(\dfrac{x-\mu^*}{{\sigma^*}}\right)^2 2 x f(x)g(\lambda x )}{\left [\dfrac{1}{{\sigma^*}}f\left(\dfrac{x-\mu^*}{{\sigma^*}}\right)+2f(x)G(\lambda x )\right]^2} dx \right \vert \\
&\leq& K  \int_{-\infty}^{\infty}  \dfrac{\dfrac{1}{{\sigma^*}^2}f\left(\dfrac{x-\mu^*}{{\sigma^*}}\right)^2 2 \vert x \vert f(x)g(\lambda x )}{\left [\dfrac{1}{{\sigma^*}}f\left(\dfrac{x-\mu^*}{{\sigma^*}}\right)\right]^2} dx \\
&\leq&  K\left\vert \int_{-\infty}^{\infty}   2 \vert x \vert  f(x)g(\lambda x ) dx \right \vert.
\end{eqnarray*}

Consider the change of variable $u=\lambda x$, for $\lambda\neq 0$, and note that $f(x) \leq f(0)$. Then,
\begin{eqnarray*}
\pi_{\MOOMIN}(\lambda) &\leq&  \dfrac{2 K f(0)}{\lambda^2} \left\vert \int_{-\infty}^{\infty}   \vert u \vert  g(u) dx \right \vert.
\end{eqnarray*}

Now, note that $\sigma^* \in [\sigma_1,\sigma_2]$, a closed finite-length interval, and similarly, $\mu^* \in [\mu_1,\mu_2]$, since the mean and variance of $2f(x)G(\lambda x )$ are bounded and finite for all $\lambda \in \R$. Then,
\begin{eqnarray*}
 \dfrac{1}{{\sigma^*}}f\left(\dfrac{x-\mu^*}{{\sigma^*}}\right)+2f(x)G(\lambda x ) \leq \dfrac{1}{{\sigma_1}}f\left(0\right)+2f(0)  = M.
\end{eqnarray*}
Let $[0,U] \subset \R_+$ be a finite-length interval. Then, using the change of variable  $u=\lambda x$, and nothing that all densities are uniformly lower bounded on $[0,U]$:
\begin{eqnarray*}
\pi_{\MOOMIN}(\lambda) &\geq& \dfrac{K}{M}\left\vert \int_{-\infty}^{\infty}  \dfrac{1}{{\sigma^*}^2}f\left(\dfrac{x-\mu^*}{{\sigma^*}}\right)^2 2 x f(x)g(\lambda x ) dx \right \vert\\
 &\geq& \dfrac{2K}{M}\left\vert \int_{0}^{U}  K_1 K_2 x g(\lambda x ) dx \right \vert\\
  &\geq& \dfrac{2K K_1 K_2}{M\lambda^2}\left\vert \int_{ 0}^{\lambda U}   u g(u ) du \right \vert.
\end{eqnarray*}
As we are interested in the tail behaviour of $\pi_{\MOOMIN}(\lambda) $, we only need to consider the case $\lambda \geq L > 0$ (using the symmetry of this function). Then, 
\begin{eqnarray*}
\pi_{\MOOMIN}(\lambda) &\geq&  \dfrac{2K K_1 K_2}{M\lambda^2}\left\vert \int_{ 0}^{L U}   u g(u ) du \right \vert,
\end{eqnarray*}
which is finite and non-zero. Consequently, the tails of the prior $\pi_{\MOOMIN}(\lambda)$ are of order $O(|\lambda|^{-2})$.

\end{itemize}

\subsection*{Proof of Proposition 2}

The proof is similar to that of \cite{johnson:2010}, adapting the different vanishing rates accordingly.

\begin{itemize}
\item[(i)] The result for local priors $\pi_{\lambda}$ follows from \cite{kass:1995} and \cite{johnson:2010}, who established the asymptotic rates for the Bayes factor estimated using the Laplace approximation 
\begin{equation*}
\widehat{BF}(1 \mid 0) = O_p(n^{- \frac{1}{2}}).
\end{equation*}
Moreover, \cite{johnson:2010} (Appendix B) showed that for non-local priors of the type $\pi_{M}(\theta)=\dfrac{\lambda^{2\tilde{k}}}{\tau_{\tilde{k}}}\pi_b(\theta)$, where $\pi_b(\theta)$ is a base prior with $2\tilde{k}$ finite moments, the Bayes factor estimated using the Laplace approximation satisfies:
\begin{equation*}
\widehat{BF}(1 \mid 0) = O_p(n^{-\tilde{k} - \frac{1}{2}}).
\end{equation*}
Noting that the approximate MOOMIN and DIMOM prior have this structure implies that the approximate Bayes factor satisfies
\begin{equation*}
\widehat{BF}(1 \mid 0) = O_p(n^{-k}),
\end{equation*}
with $k = 3/2$ for the DIMOM prior, and $k = 5/2$ for the approximate MOOMIN prior.
\item[(ii)] The result follows from \cite{kass:1995} (see also the discussions in \citep{johnson:2010}). In our notation,  \cite{kass:1995} showed that the Bayes factors, approximated using the Laplace approximation and under standard regularity conditions (satisfied by skew-symmetric models), in favour of the alternative satisfies
\begin{equation*}
\frac{1}{n}\log \widehat{BF}(1\mid 0)
\;\xrightarrow{\Pr}\;
\mathrm{KL}\!\left(
f(\cdot \mid \mu^\star,\sigma^\star)\,\|\, s_{f,G}( \cdot \mid \mu_0,\sigma_0, \lambda_0)
\right),
\end{equation*}
where $\mathrm{KL}(\cdot\|\cdot)$ denotes the Kullback--Leibler divergence, and $(\mu^\star,\sigma^\star)$ are the values of the location and scale parameters that minimise the Kullback--Liebler divergence to the true generating model $s_{f,G}( \cdot \mid \mu_0,\sigma_0, \lambda_0)$.
Equivalently,
\begin{equation*}
\widehat{BF}(1\mid 0)
=
\exp\!\left\{
n\,\mathrm{KL}\!\left(
f(\cdot \mid \mu^\star,\sigma^\star)\,\|\, s_{f,G}( \cdot \mid \mu_0,\sigma_0, \lambda_0)
\right)
+ o_p(n)
\right\}.
\end{equation*}
The constant $c_{\btheta_0}$ is thus the Kullback-Leibler divergence $\mathrm{KL}\!\left(f(\cdot \mid \mu^\star,\sigma^\star)\,\|\, s_{f,G}( \cdot \mid \mu_0,\sigma_0, \lambda_0)\right)$.
\end{itemize}

\subsection*{Simulation results}

\begin{figure}[h!]
    \centering
\begin{tabular}{c c c}
\includegraphics[width=0.3\textwidth]{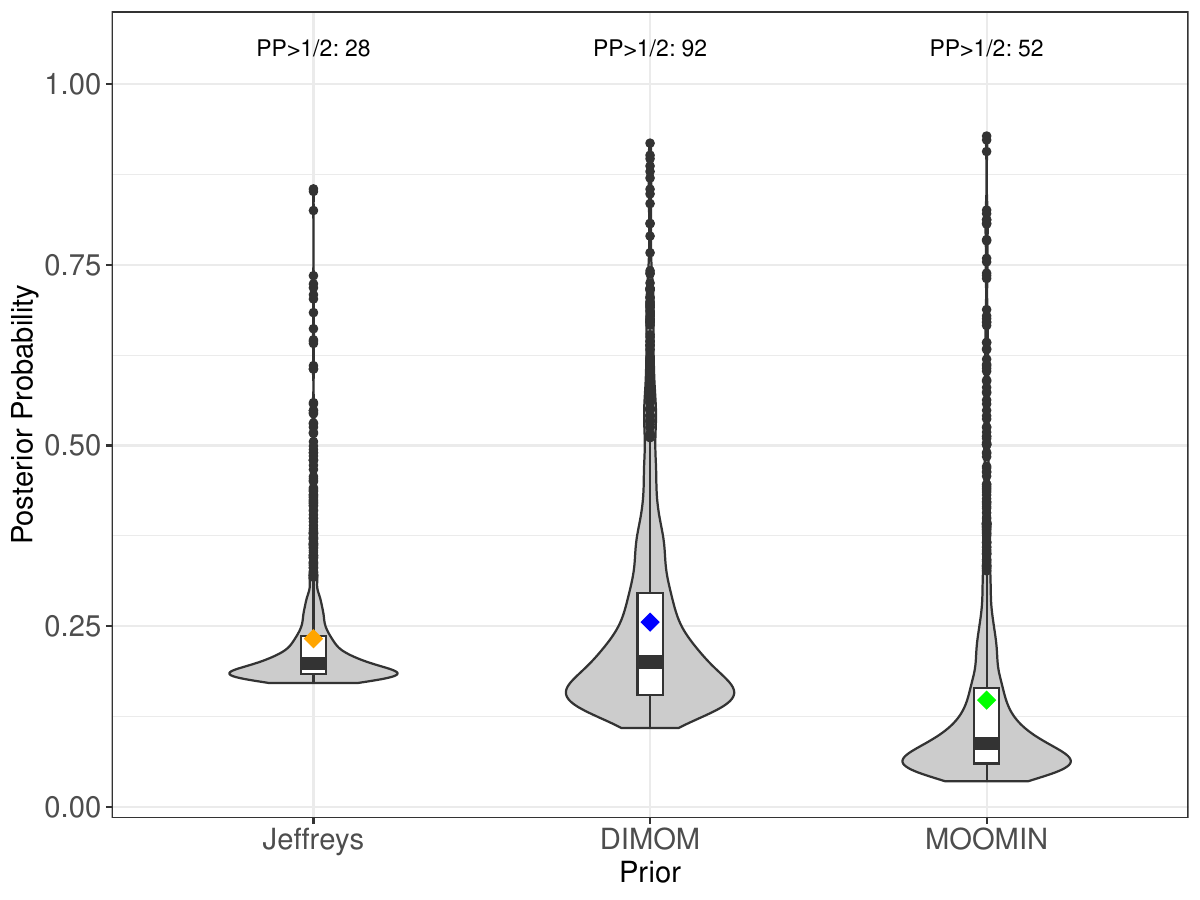} & 
\includegraphics[width=0.3\textwidth]{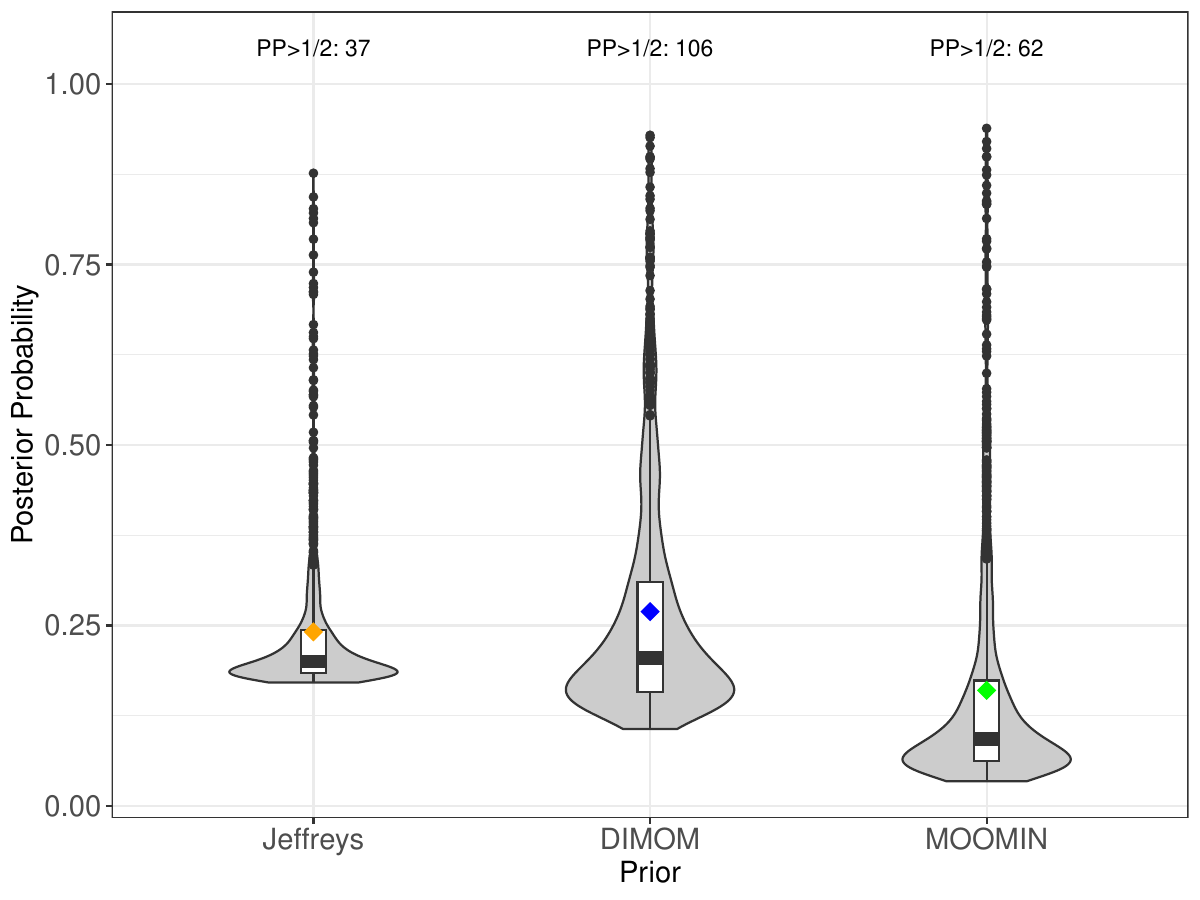} & 
\includegraphics[width=0.3\textwidth]{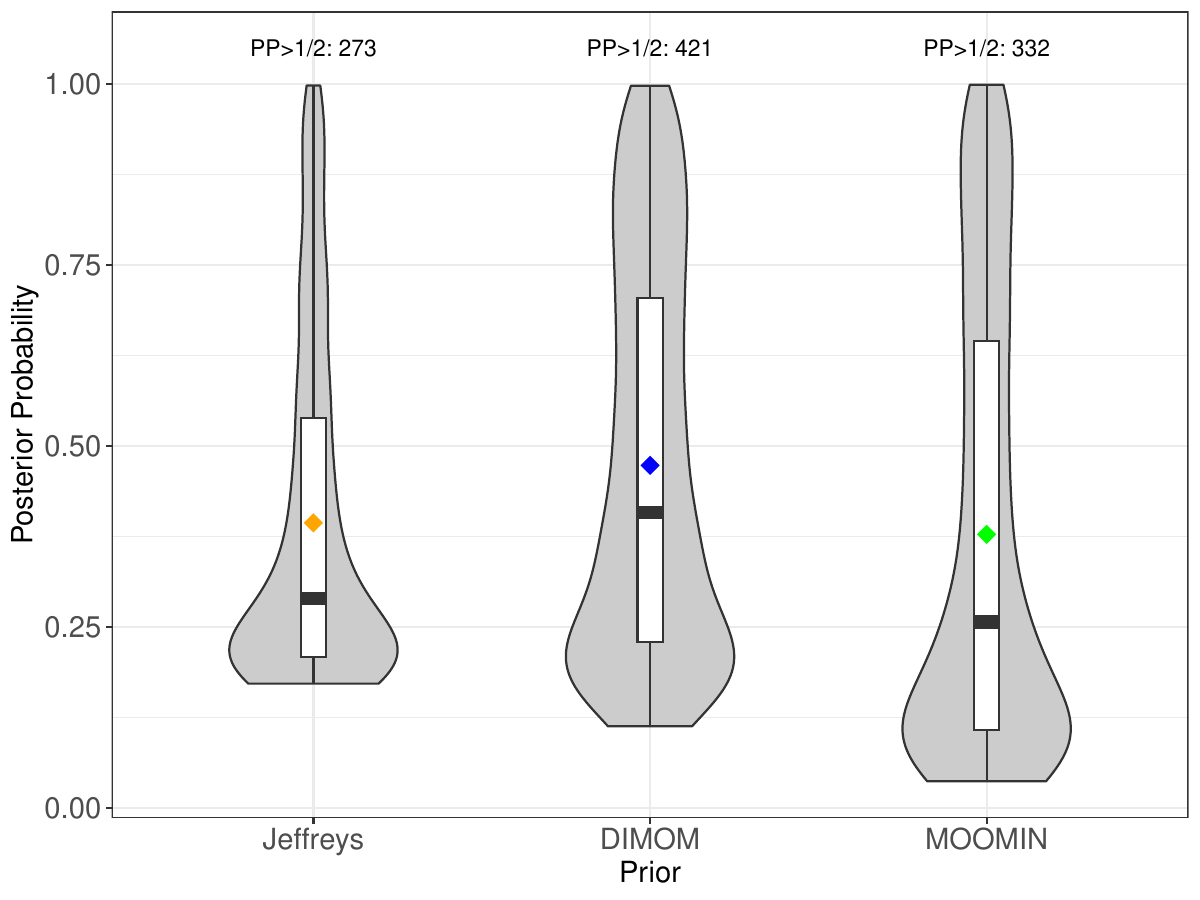} \\
 (a) & (b) & (c)\\
 \end{tabular}
    \caption{Simulation results for $n=50$: (a) $\lambda = 0$, (b) $\lambda = 1$, and (c) $\lambda = 2.5$.}
    \label{fig:sn_sim_n50}
\end{figure}

\begin{figure}[h!]
    \centering
\begin{tabular}{c c c}
\includegraphics[width=0.3\textwidth]{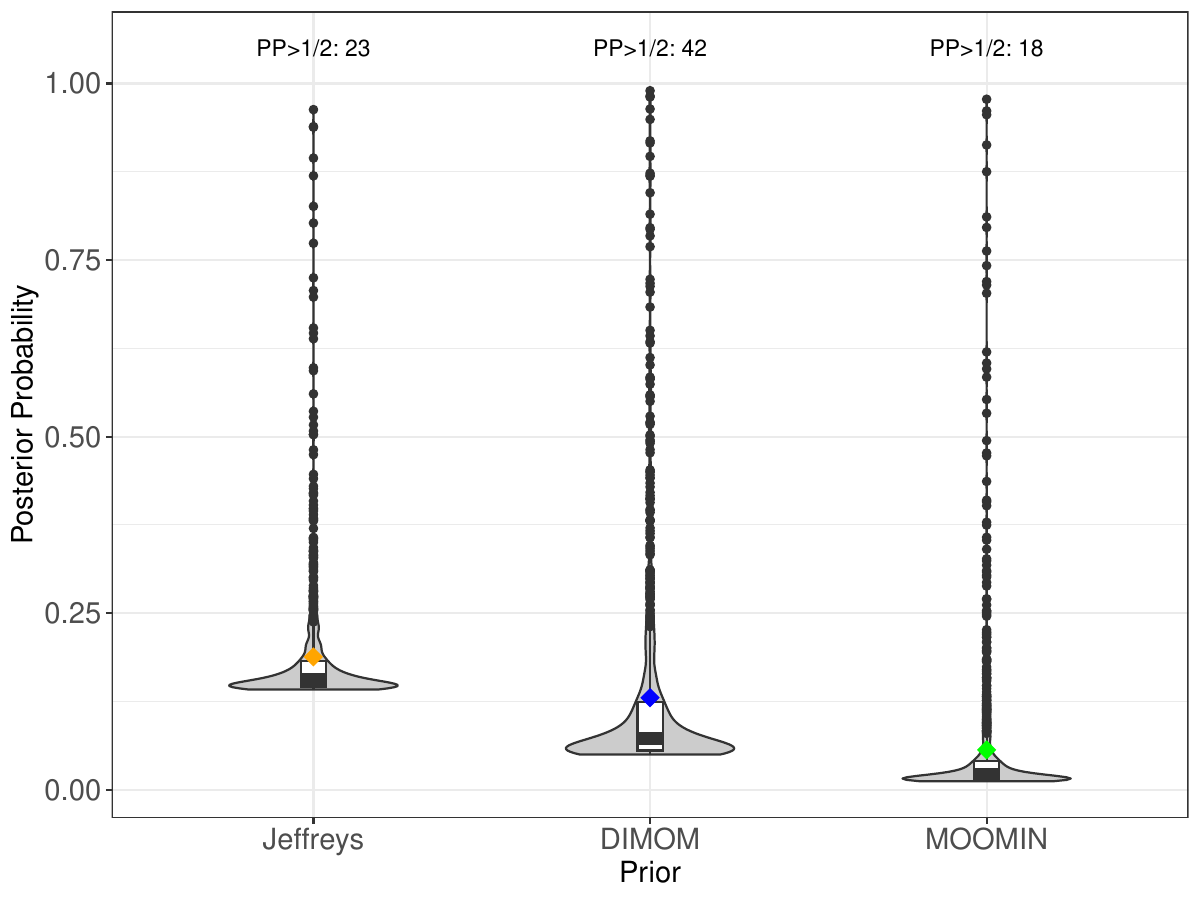} & 
\includegraphics[width=0.3\textwidth]{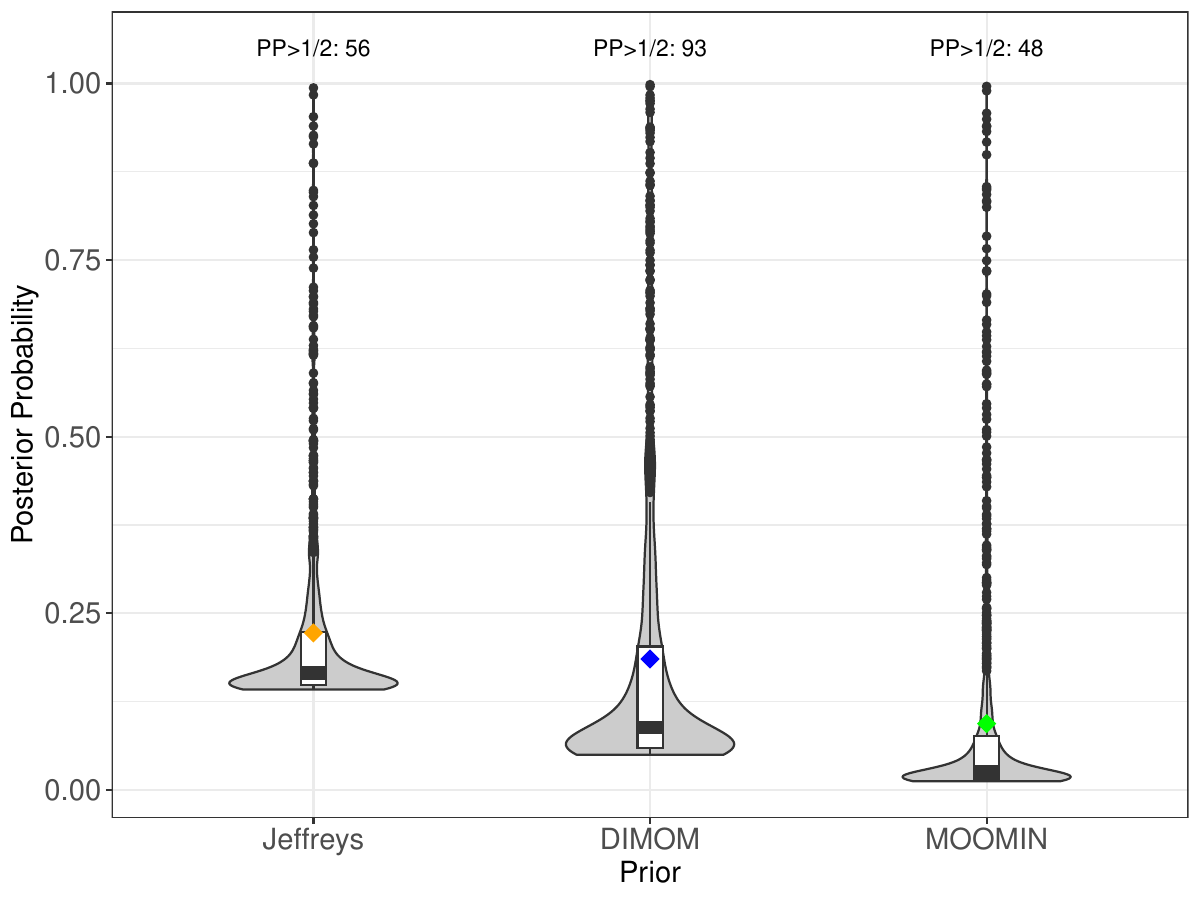} & 
\includegraphics[width=0.3\textwidth]{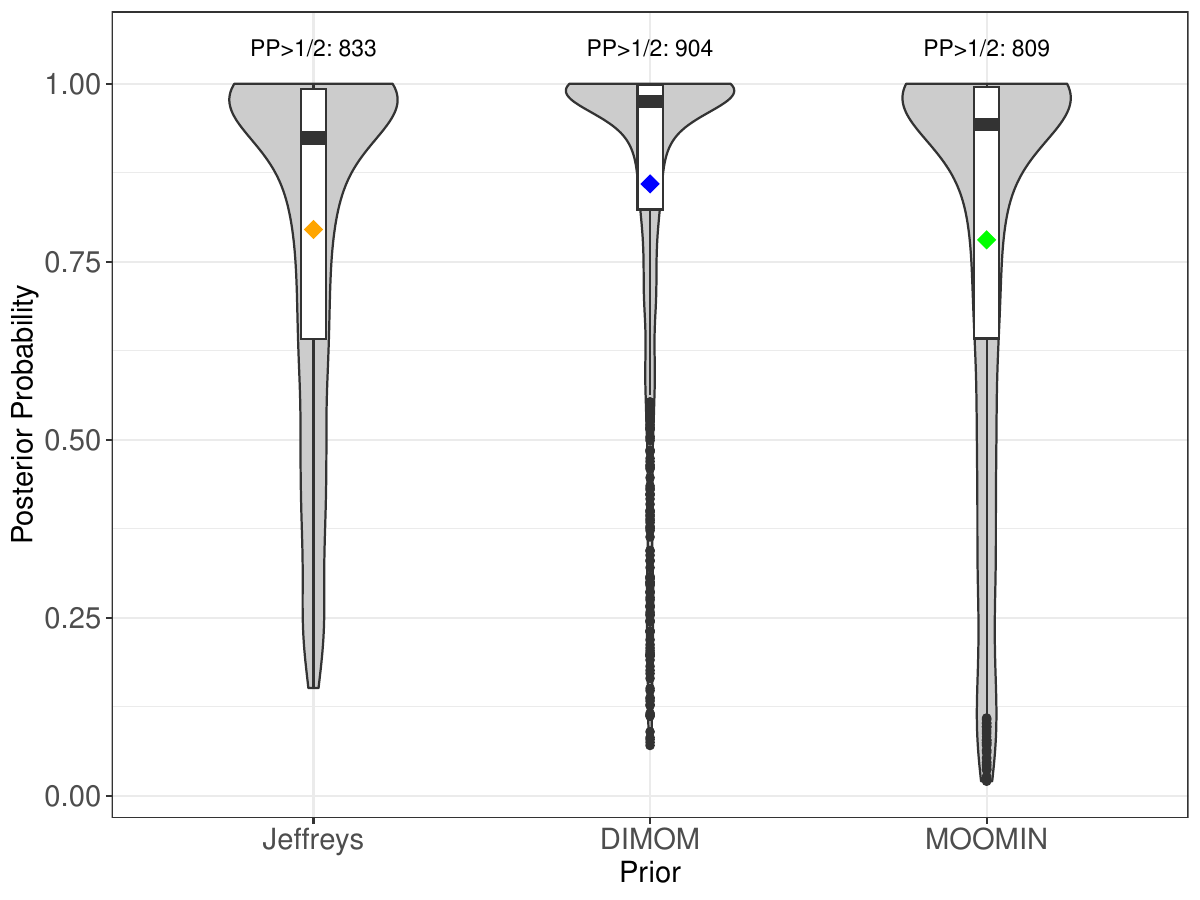} \\
 (a) & (b) & (c)\\
 \end{tabular}
    \caption{Simulation results for $n=200$: (a) $\lambda = 0$, (b) $\lambda = 1$, and (c) $\lambda = 2.5$.}
    \label{fig:sn_sim_n200}
\end{figure}

\begin{figure}[h!]
    \centering
\begin{tabular}{c c c}
\includegraphics[width=0.3\textwidth]{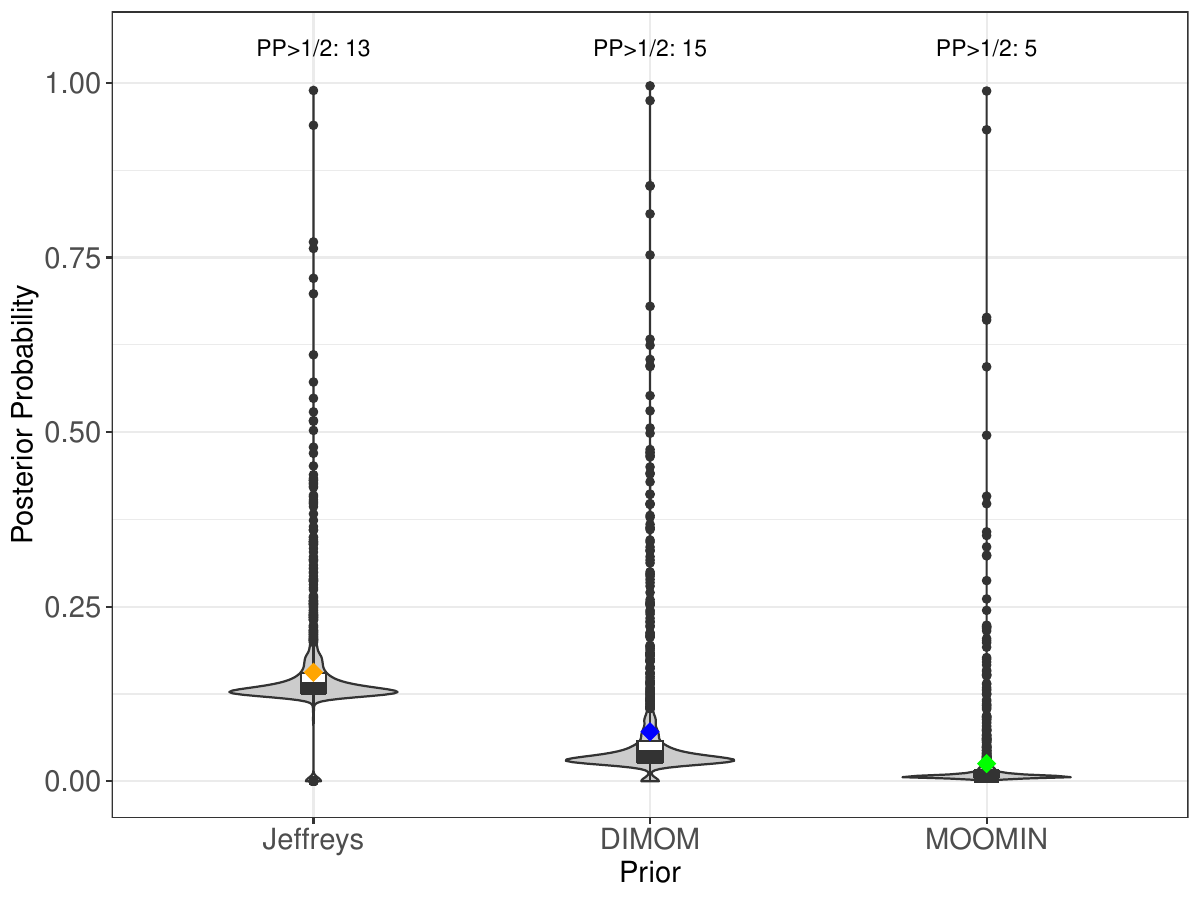} & 
\includegraphics[width=0.3\textwidth]{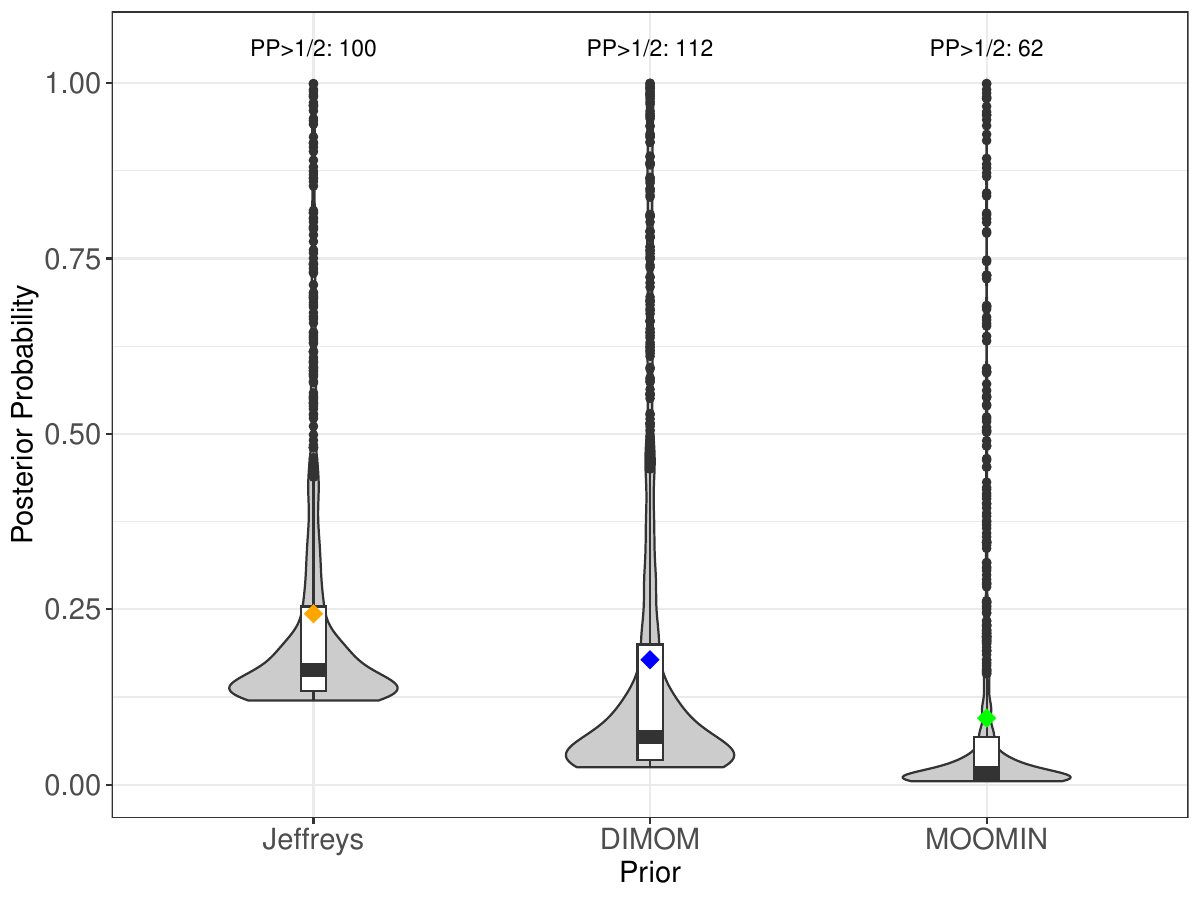} & 
\includegraphics[width=0.3\textwidth]{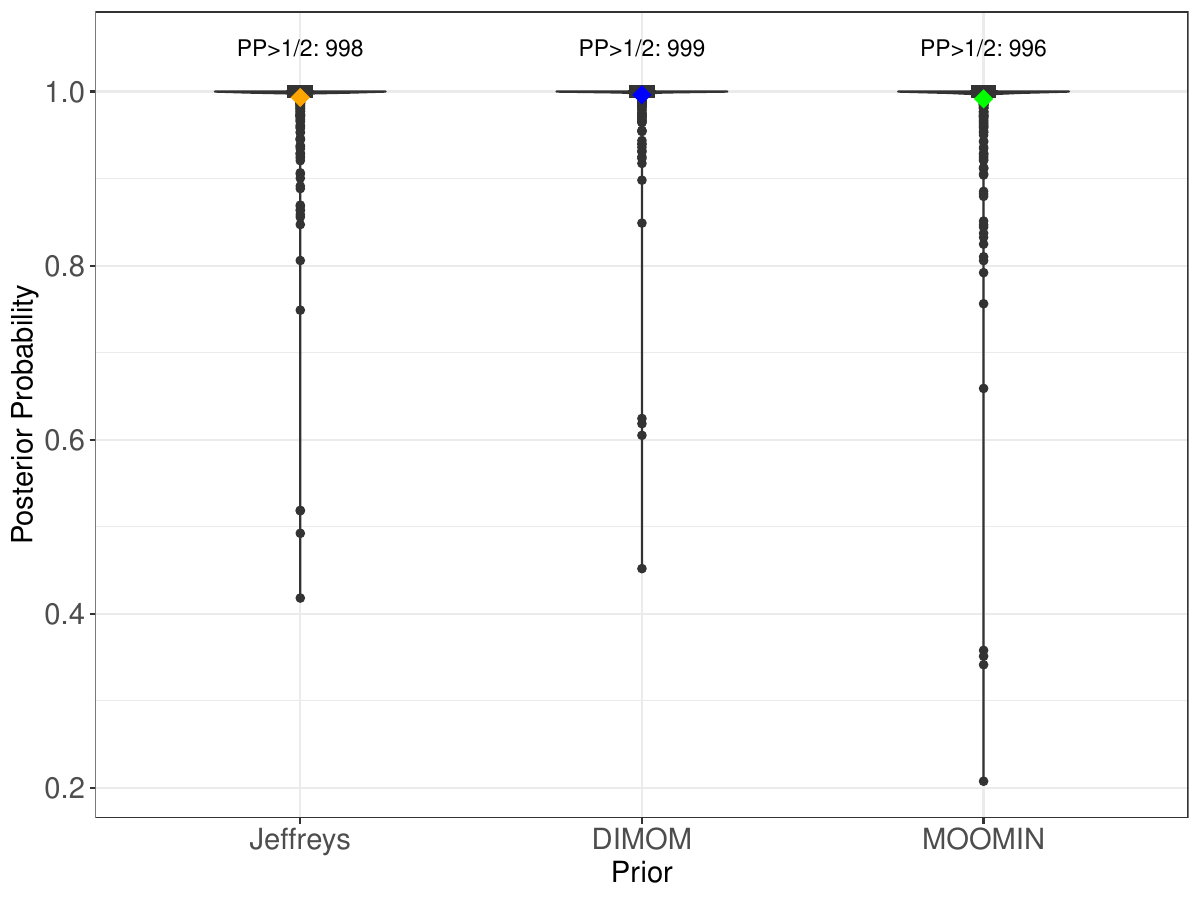} \\
 (a) & (b) & (c)\\
 \end{tabular}
    \caption{Simulation results for $n=500$: (a) $\lambda = 0$, (b) $\lambda = 1$, and (c) $\lambda = 2.5$.}
    \label{fig:sn_sim_n500}
\end{figure}

\clearpage
{
\subsection*{MOOMIN prior for other distributions}

\subsubsection*{Skew-symmetric distributions}

We now analyse the discrepancy measure, signed discrepancy measure and the MOOMIN prior for two additional skew-symmetric models:

\begin{enumerate}
    \item Skew-logistic distribution, where the baseline is the logistic distribution
    \begin{align*}
        f(x) &= \dfrac{e^{-x}}{(1+e^{-x})^2},\\
        G(x) &= \dfrac{1}{1+e^{-x}}.
    \end{align*}
        \item Skew-sech distribution, where the baseline is the hyperbolic secant distribution
    \begin{align*}
        f(x) &= \dfrac{1}{2}\operatorname{sech}\left(\dfrac{\pi}{2}x\right),\\
        G(x) &= {\frac {2}{\pi }}\arctan \!\left[\exp \!\left({\frac {\pi }{2}}\,x\right)\right].
    \end{align*}
\end{enumerate}

The corresponding plots are shown in Figures \ref{fig:slogis_min} and \ref{fig:shs_min}. We observe the following:
\begin{itemize}
    \item In the skew-normal case presented in the main manuscript, the parameter $\lambda$ has a ``slow'' effect, in the sense that the distribution is virtually symmetric for $\lvert \lambda \rvert \leq 1$ and skewness is induced only gradually. This leads to a MOOMIN prior that heavily penalises values in this region and exhibits polynomial tails.
    \item The skew-logistic case represents an intermediate scenario, in which the parameter $\lambda$ has a ``faster'' effect, resulting in less prior mass being removed around $\lambda = 0$.
    \item In the skew-sech case, the parameter $\lambda$ has a ``fast'' effect, as the shape of the distribution changes rapidly as $\lambda$ moves away from zero. Consequently, the corresponding MOOMIN prior penalises only a small neighbourhood around zero.
\end{itemize}

\begin{figure}[h!]
    \centering
\begin{tabular}{c c c}
\includegraphics[width=0.3\textwidth]{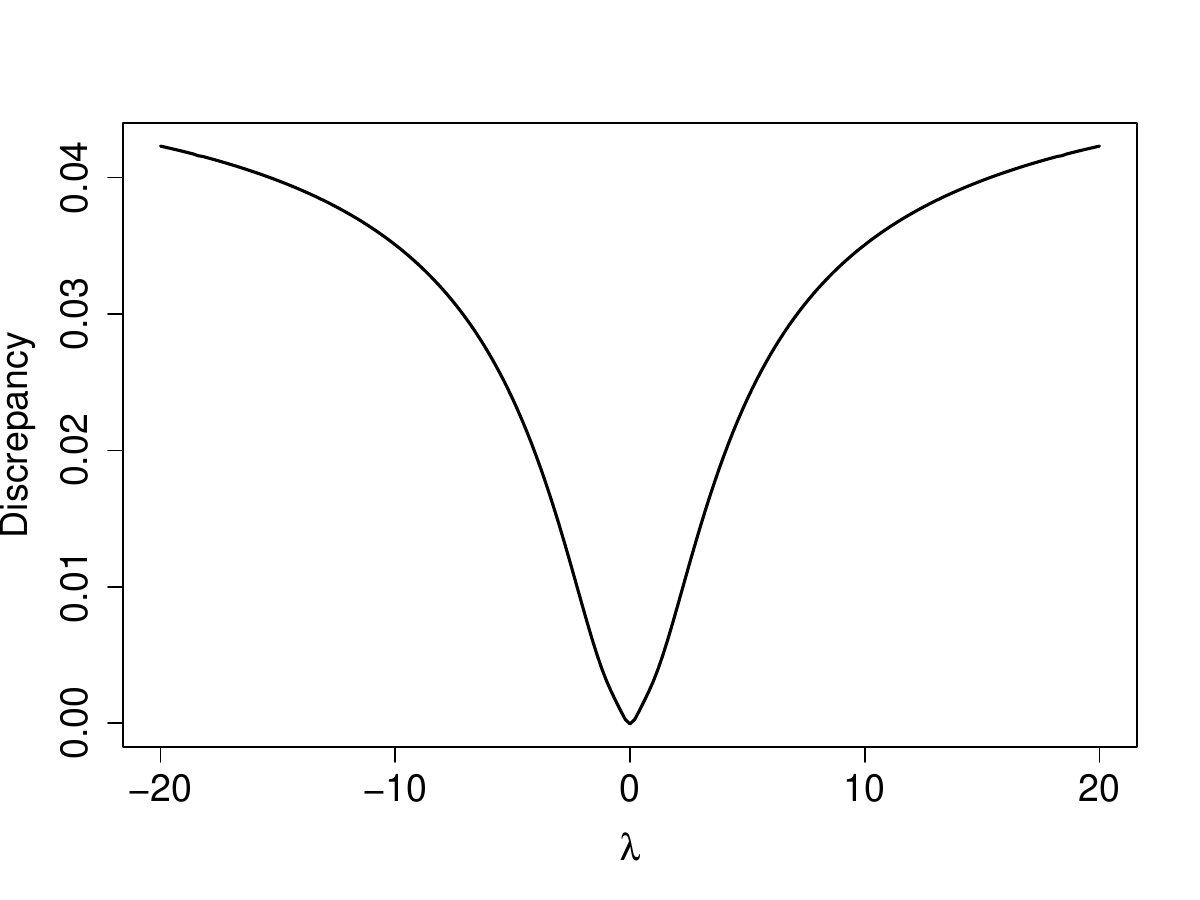} & 
\includegraphics[width=0.3\textwidth]{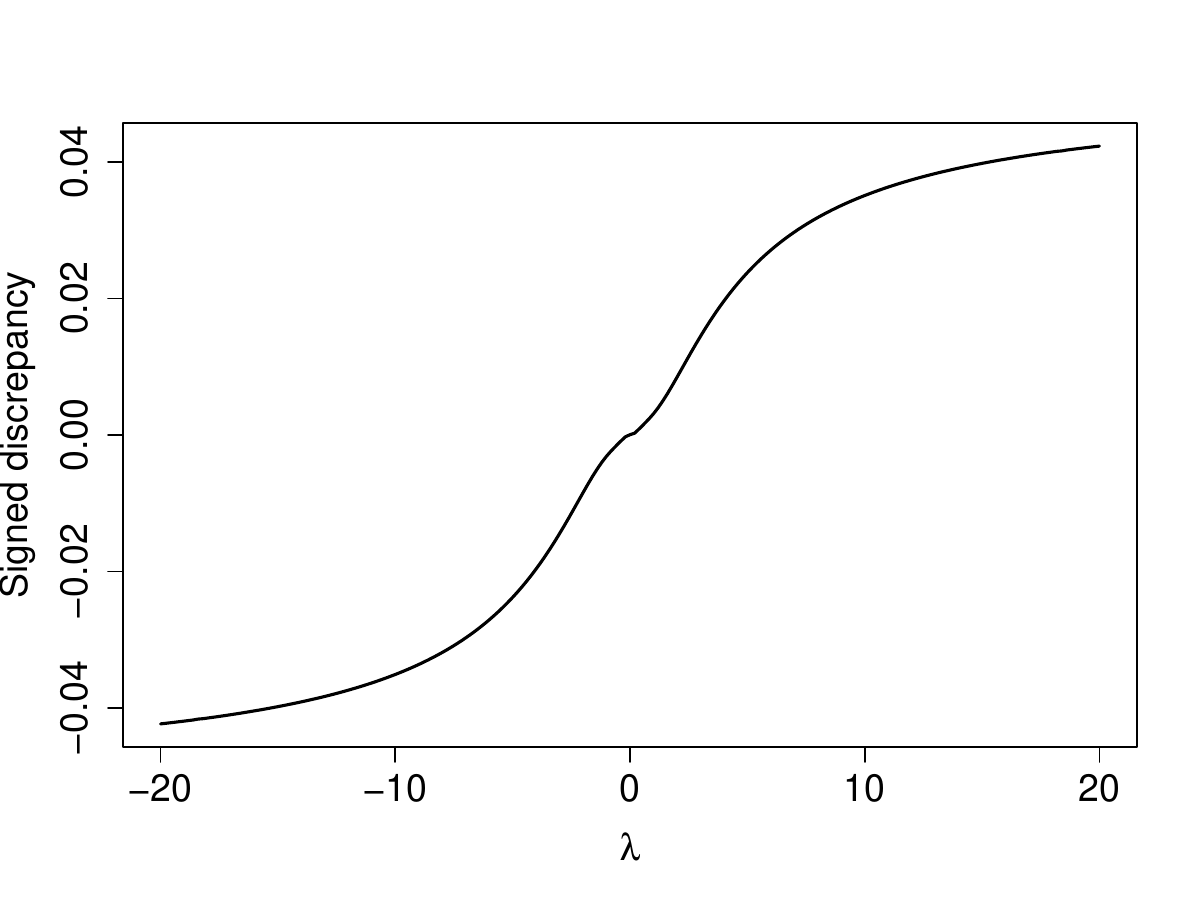} & 
\includegraphics[width=0.3\textwidth]{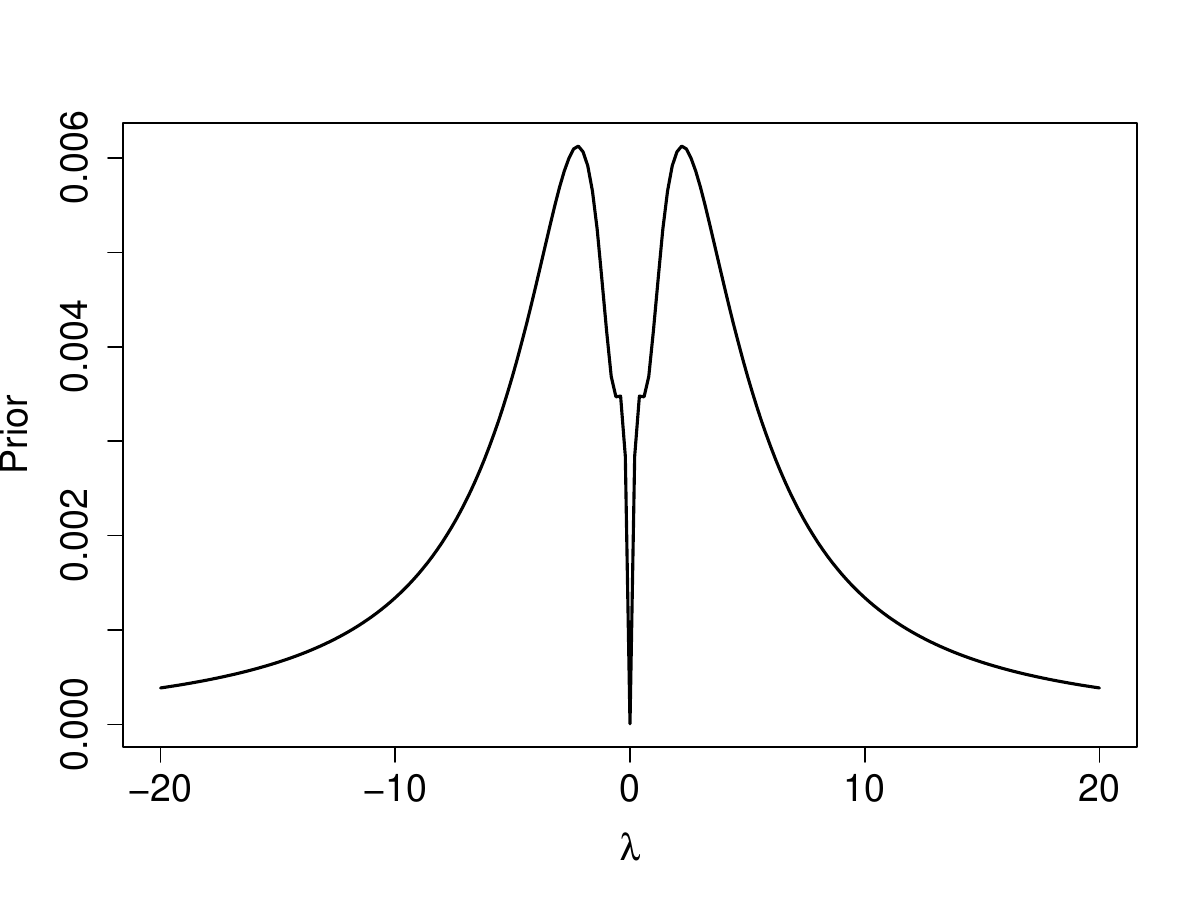} \\
 (a) & (b) & (c)\\
 \end{tabular}
    \caption{Skew-logistic distribution: (a) Discrepancy measure, (b) signed discrepancy measure, and (c) MOOMIN prior.}
    \label{fig:slogis_min}
\end{figure}

\begin{figure}[h!]
    \centering
\begin{tabular}{c c c}
\includegraphics[width=0.3\textwidth]{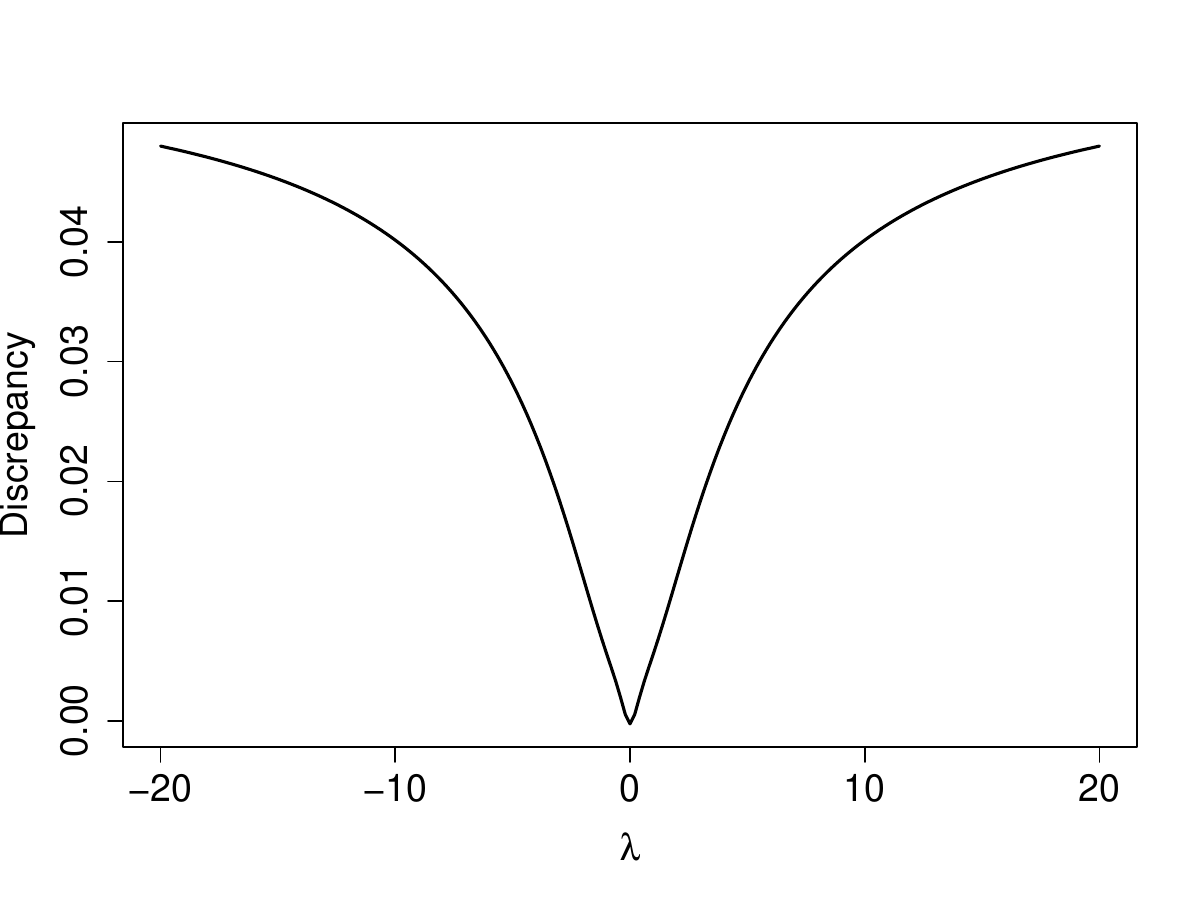} & 
\includegraphics[width=0.3\textwidth]{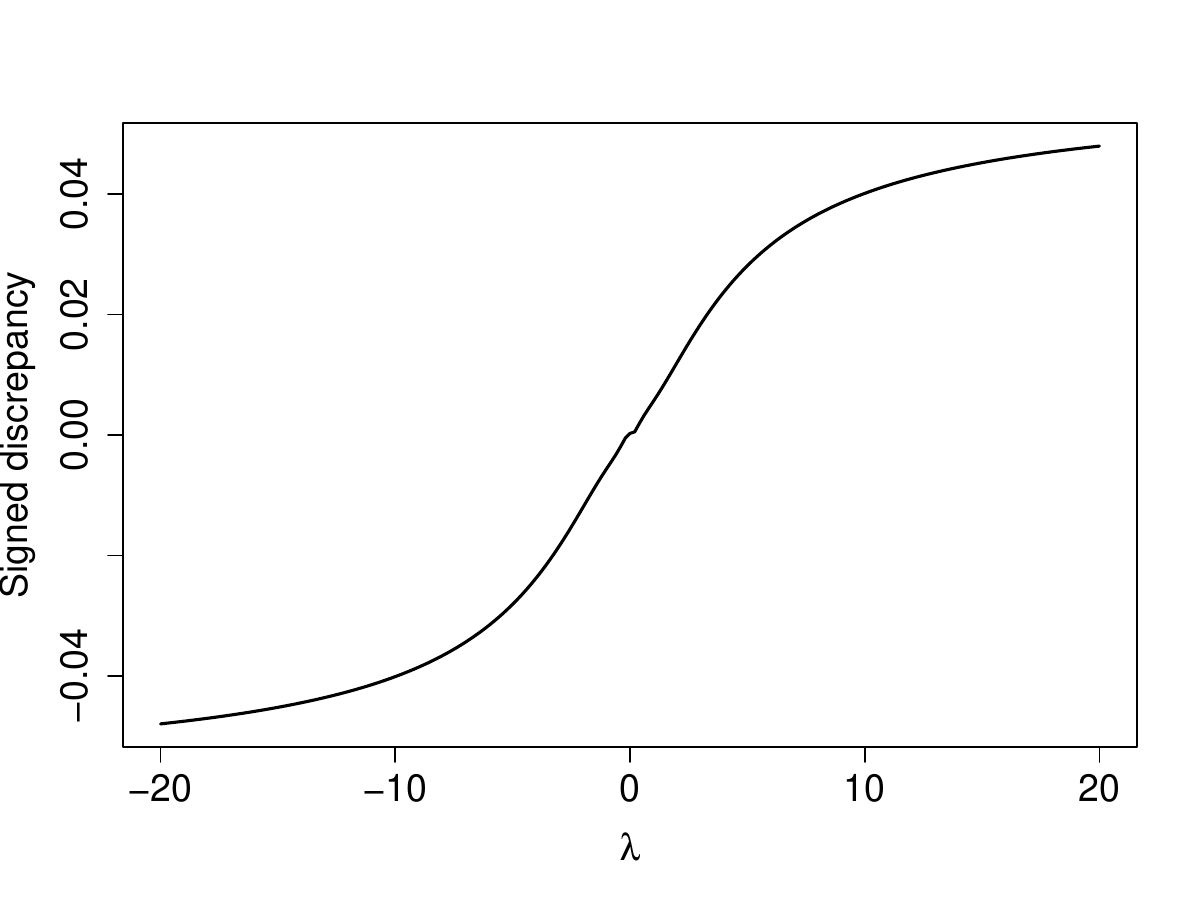} & 
\includegraphics[width=0.3\textwidth]{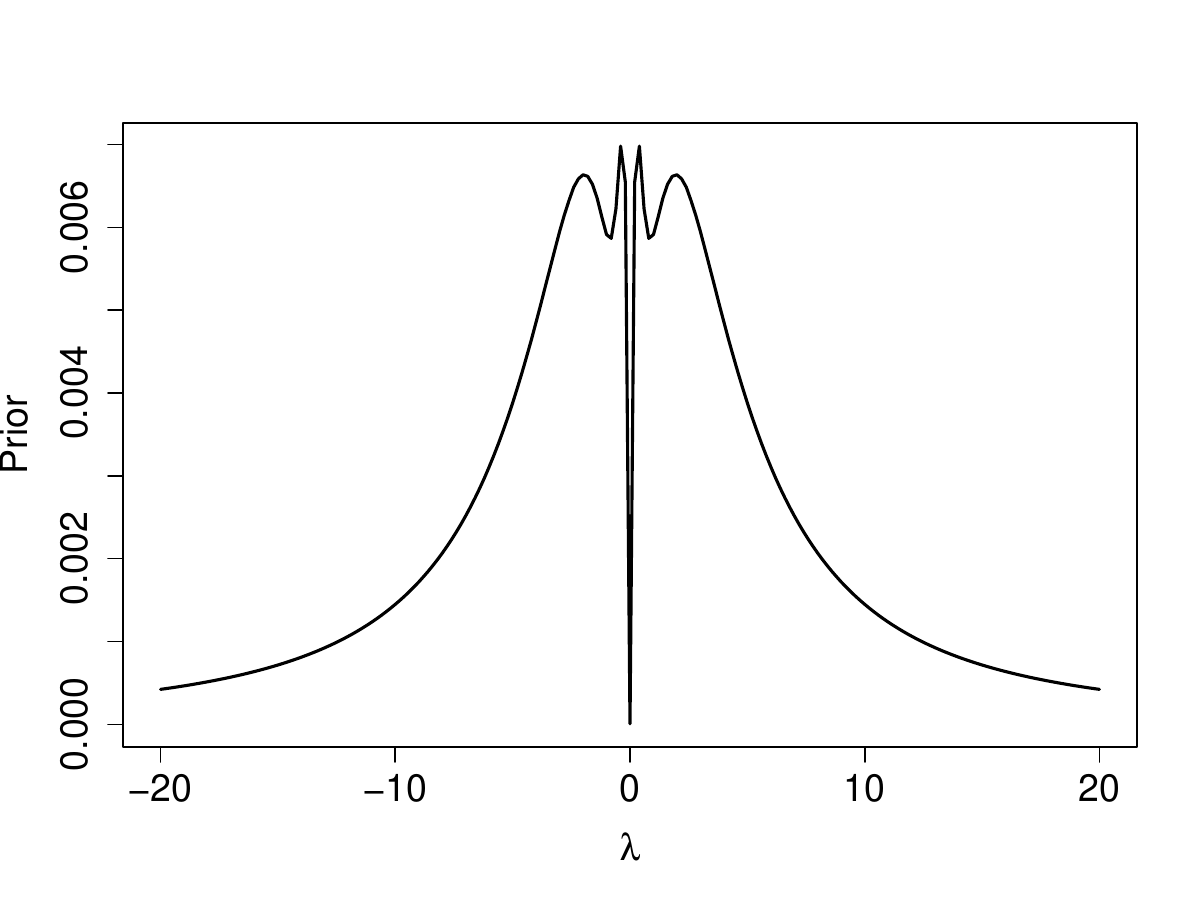} \\
 (a) & (b) & (c)\\
 \end{tabular}
    \caption{Skew-sech distribution: (a) Discrepancy measure, (b) signed discrepancy measure, and (c) MOOMIN prior.}
    \label{fig:shs_min}
\end{figure}

\pagebreak

\subsubsection*{Two-piece distributions}
In this section, we analyse the discrepancy measure, the signed discrepancy measure, and the MOOMIN prior for three members of the family of two-piece distributions \citep{rubio:2020}. The two-piece family is constructed by continuously joining two half-density functions. Several parameterisations of this family exist \citep{rubio:2014}; here, we focus on the parameterisation given below. 
\begin{align*}
s(x \mid \mu,\sigma,\epsilon) = 
\begin{cases}
\dfrac{2}{\sigma_1 + \sigma_2} f\left(\dfrac{x-\mu}{\sigma_1}\right) & \text{if } x< \mu,\\
\dfrac{2}{\sigma_1 + \sigma_2} f\left(\dfrac{x-\mu}{\sigma_2}\right) & \text{if } x \geq \mu,
\end{cases}
\end{align*}
where $f$ denotes a continuous symmetric (about $x=0$) probability density function (pdf) with support on ${\mathbb R}$, $\sigma_1 = \sigma(1+\tanh(\epsilon))$, $\sigma_2 = \sigma(1-\tanh(\epsilon))$, $\mu\in\R$, $\sigma>0$, and $\epsilon\in\R$. We now define 
\begin{eqnarray*}
\D_{\min}(\epsilon) = \inf_{\mu,\sigma} \int_{-\infty}^{\infty}  \frac{\frac{1}{\sigma^2}f\left(\dfrac{x-\mu}{\sigma}\right)^2}{\frac{1}{\sigma}f\left(\frac{x-\mu}{\sigma}\right)+ s(x \mid 0,1,\epsilon)} \, dx,  
\end{eqnarray*}
and analyse the induced MOOMIN prior on $\epsilon$.
We consider the cases where $f$ is the standard normal, the logistic, and the $\operatorname{sech}$ density function. The corresponding plots are shown in Figures \ref{fig:tpn_min}-\ref{fig:tphs_min}. These figures show that the parameter $\epsilon$ plays a very similar role in all cases. This effect is inherent to the construction of the family of two-piece distributions: $\epsilon$ only affects the scale and the mass allocated to each side of the mode, without altering other features such as tail weight, the mode itself, or kurtosis.
\begin{figure}[h!]
    \centering
\begin{tabular}{c c c}
\includegraphics[width=0.3\textwidth]{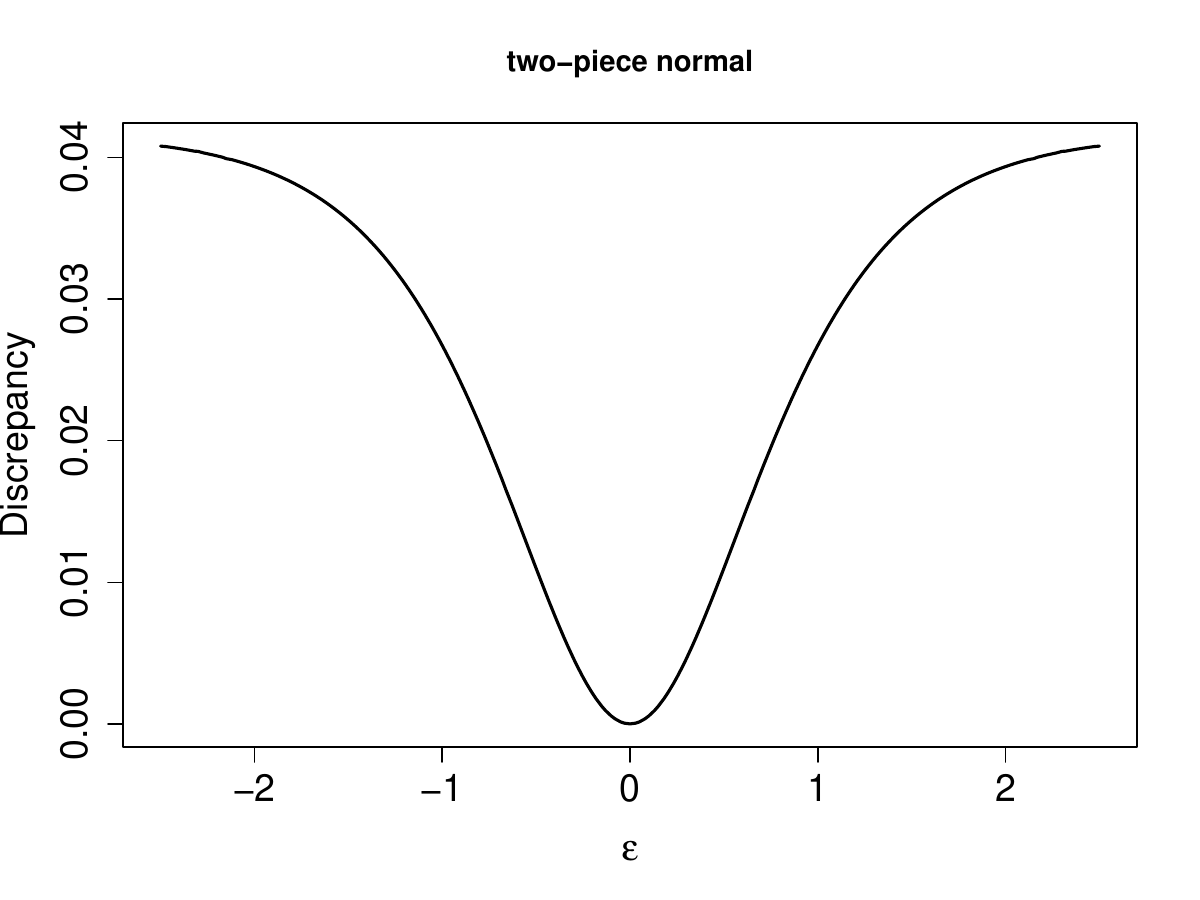} & 
\includegraphics[width=0.3\textwidth]{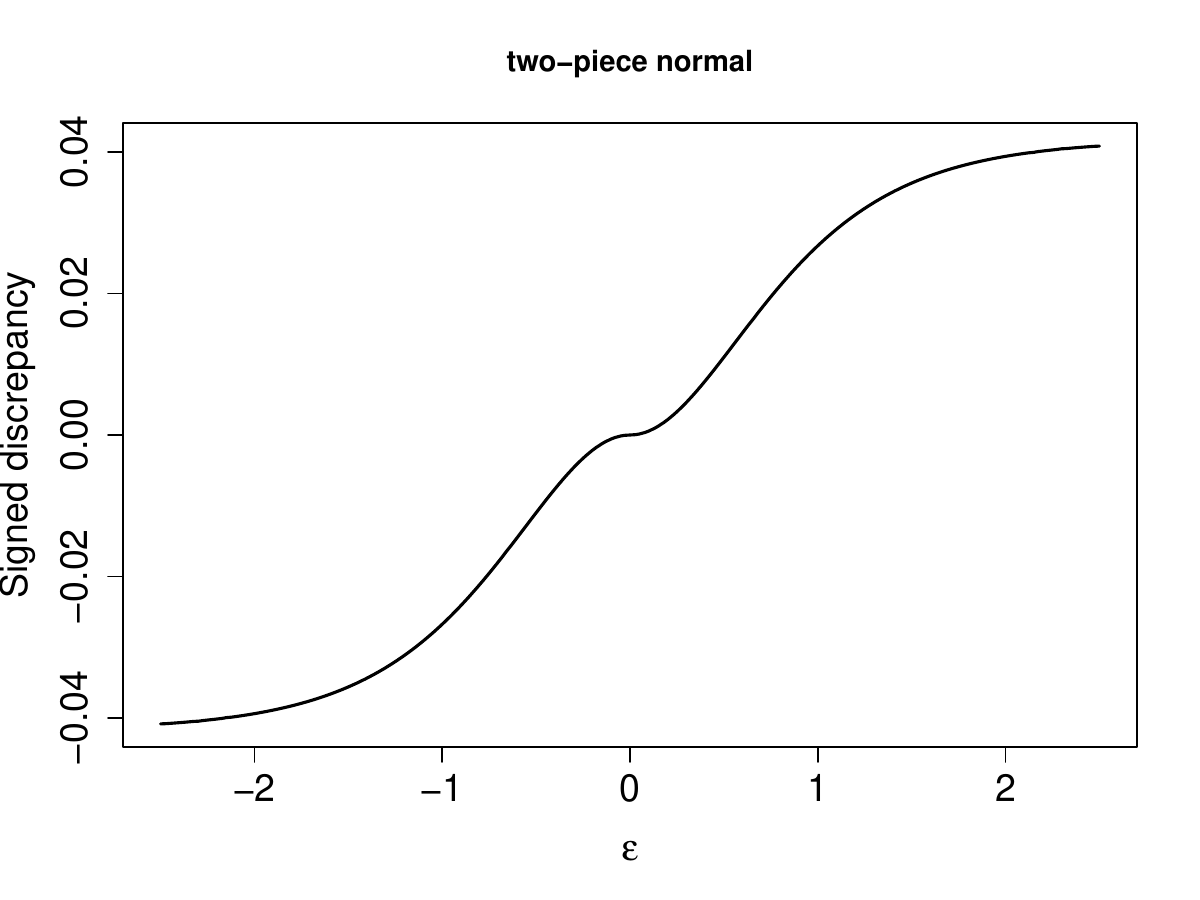} & 
\includegraphics[width=0.3\textwidth]{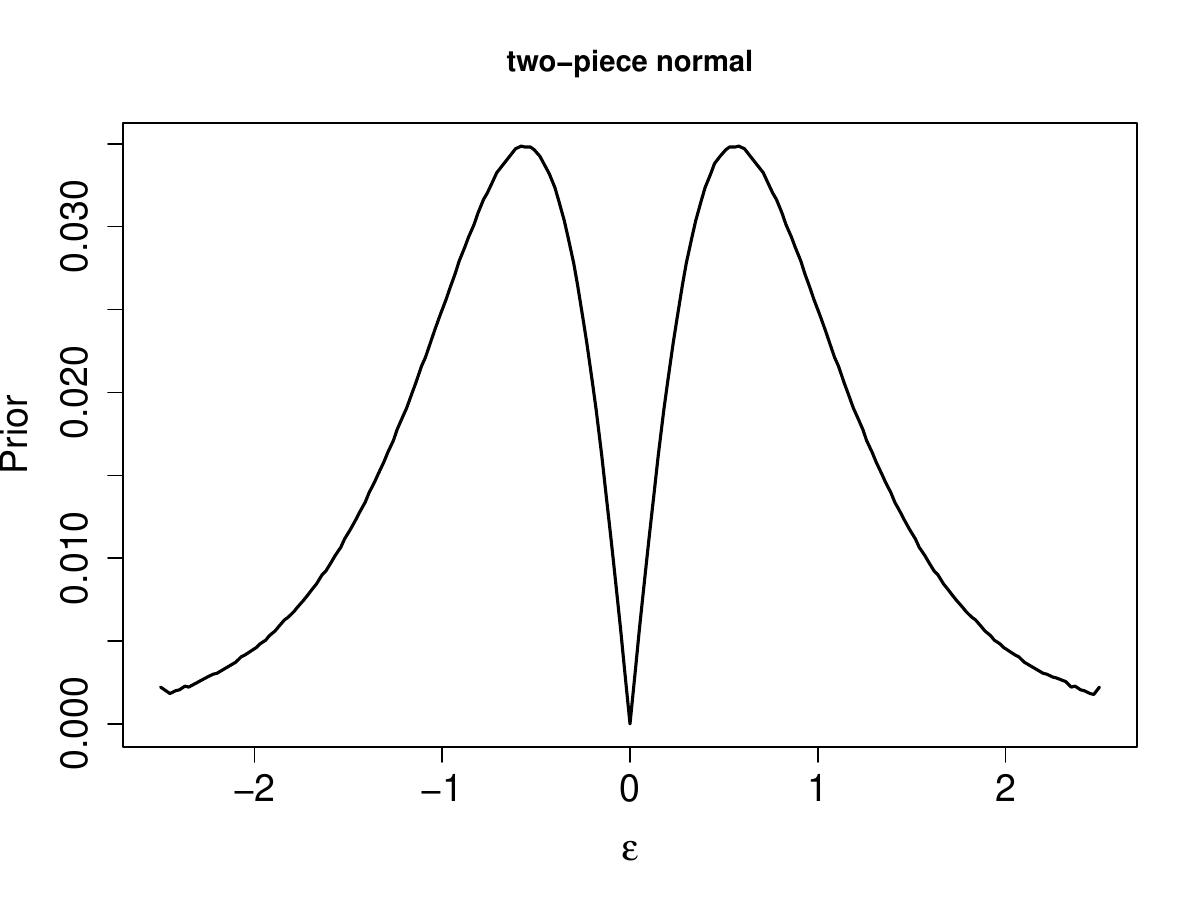} \\
 (a) & (b) & (c)\\
 \end{tabular}
    \caption{Two-piece normal distribution: (a) Discrepancy measure, (b) signed discrepancy measure, and (c) MOOMIN prior.}
    \label{fig:tpn_min}
\end{figure}

\begin{figure}[h!]
    \centering
\begin{tabular}{c c c}
\includegraphics[width=0.3\textwidth]{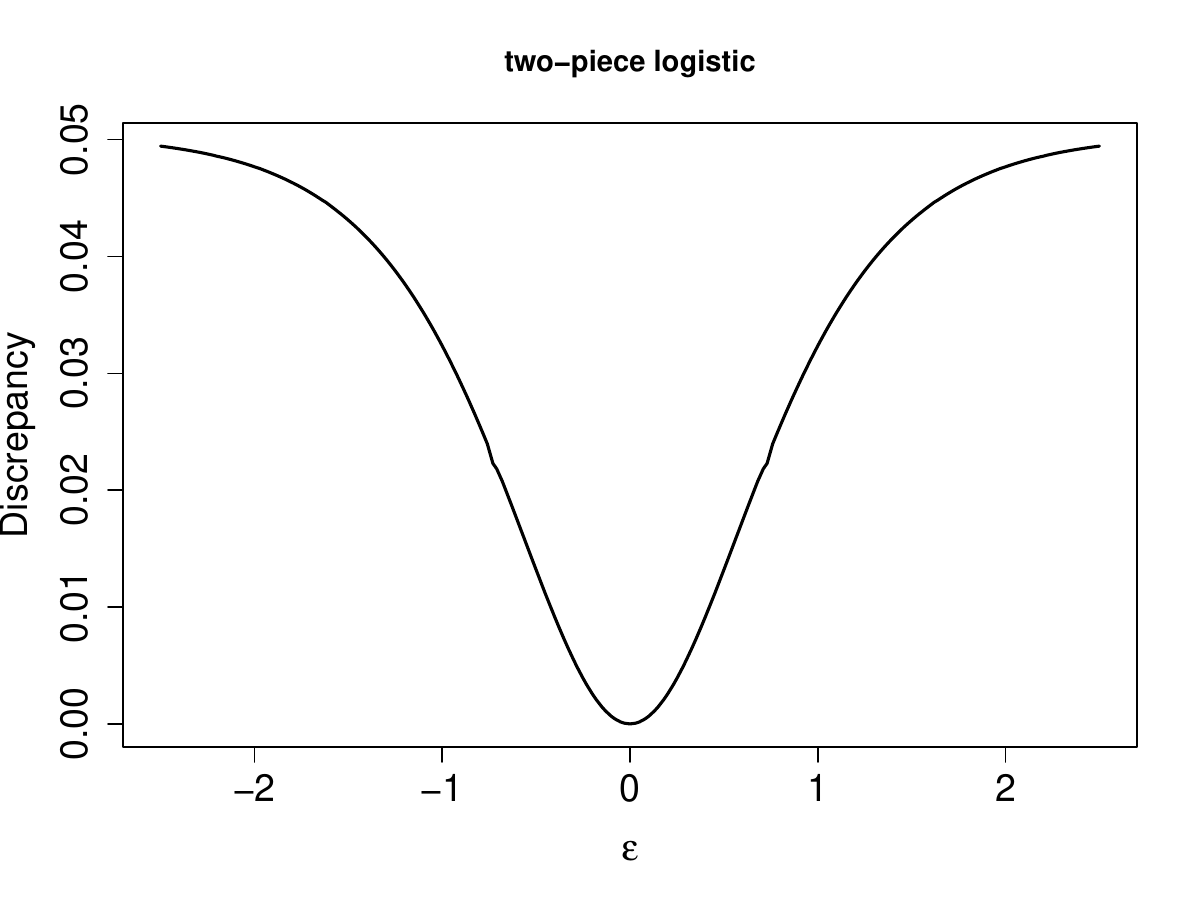} & 
\includegraphics[width=0.3\textwidth]{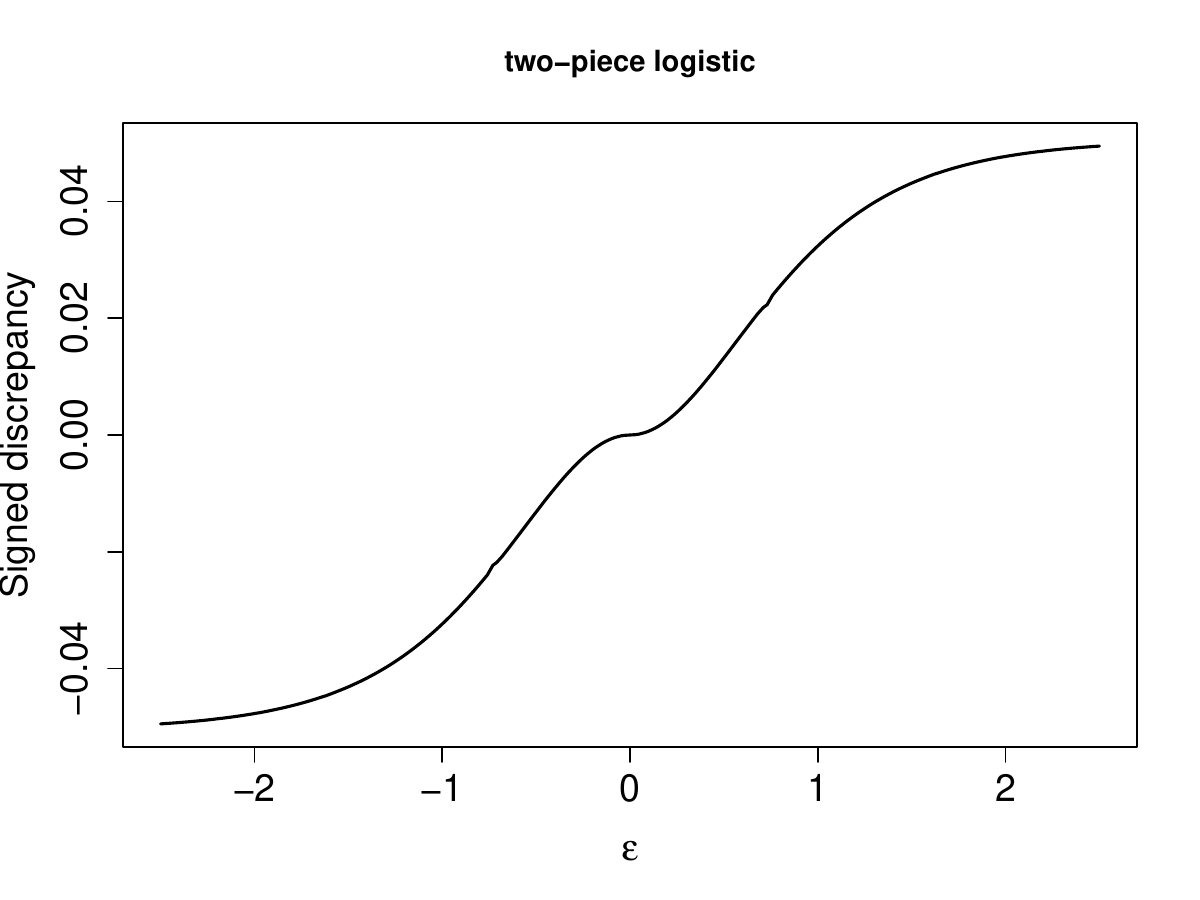} & 
\includegraphics[width=0.3\textwidth]{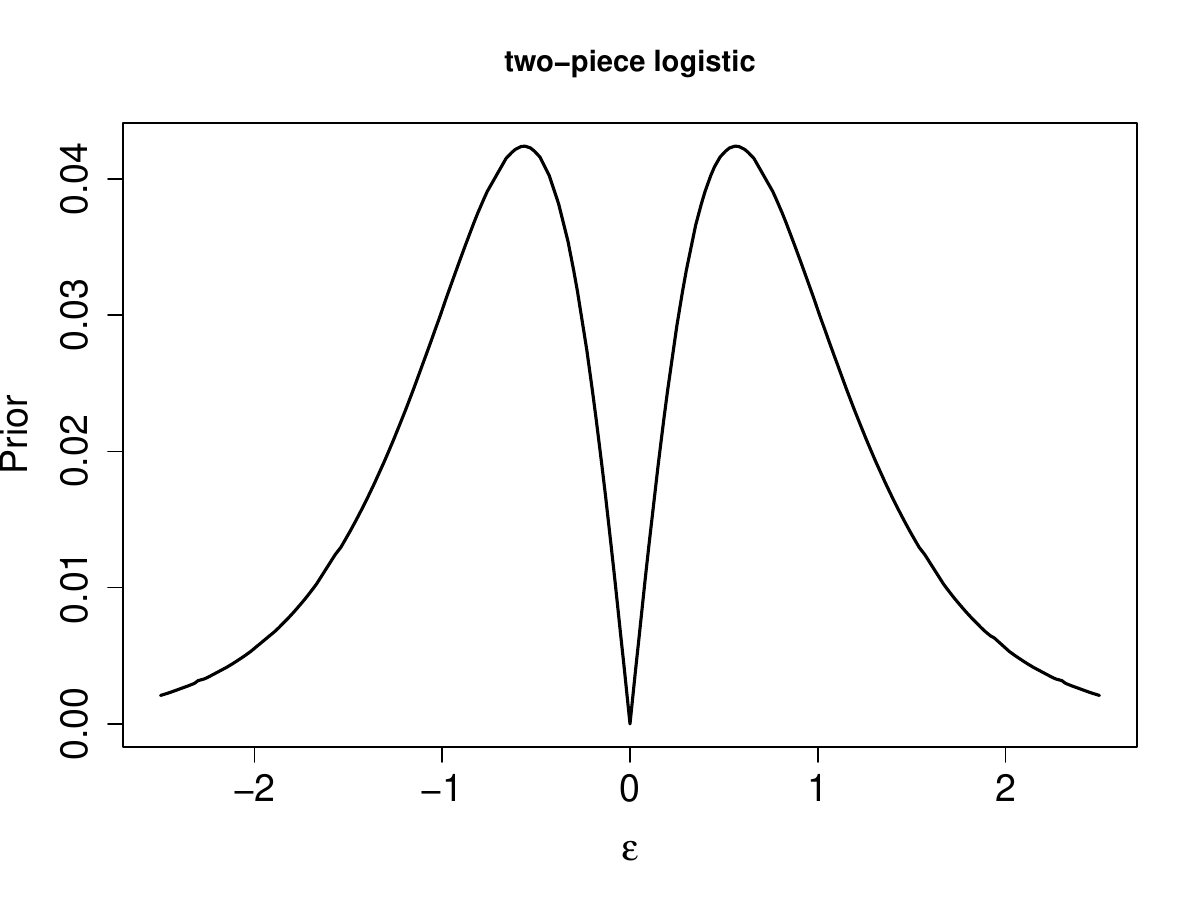} \\
 (a) & (b) & (c)\\
 \end{tabular}
    \caption{Two-piece logistic distribution: (a) Discrepancy measure, (b) signed discrepancy measure, and (c) MOOMIN prior.}
    \label{fig:tplogis_min}
\end{figure}

\begin{figure}[h!]
    \centering
\begin{tabular}{c c c}
\includegraphics[width=0.3\textwidth]{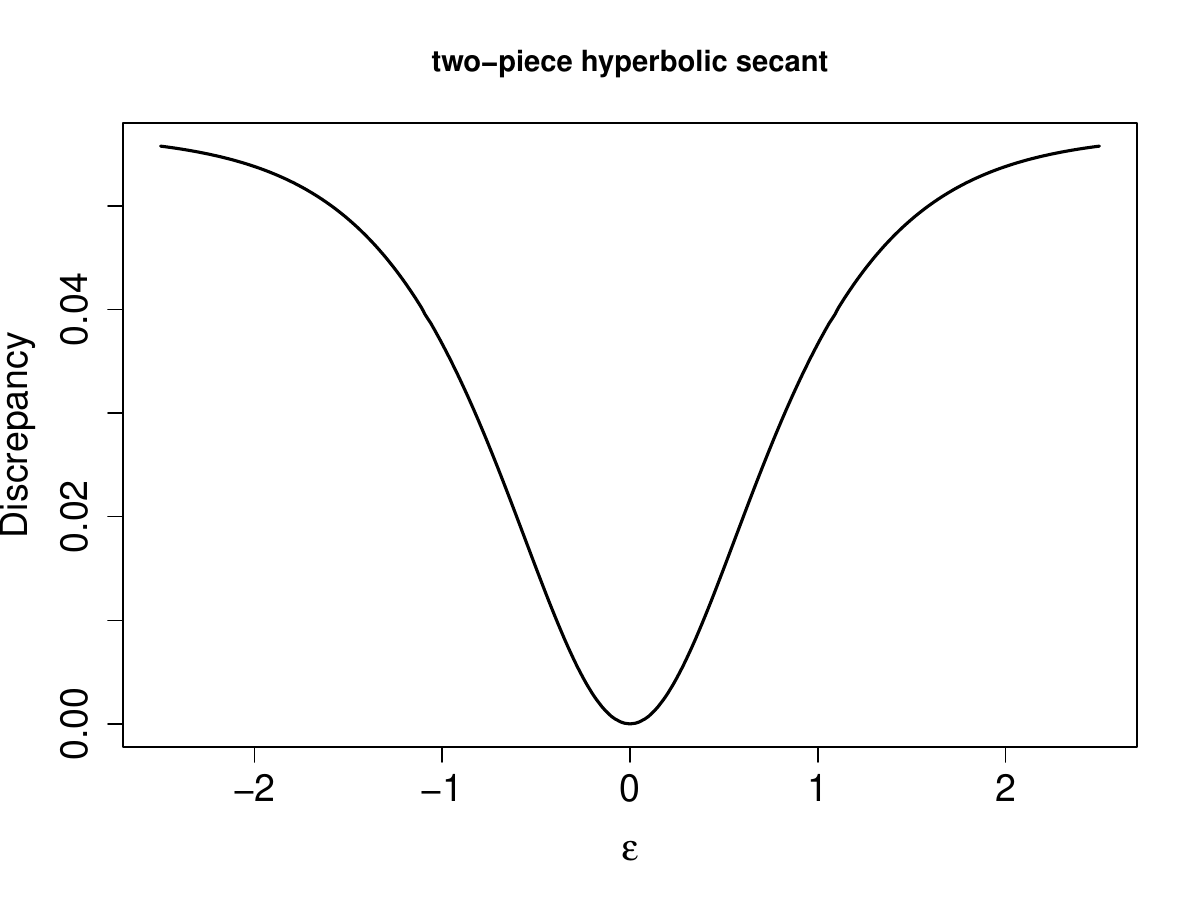} & 
\includegraphics[width=0.3\textwidth]{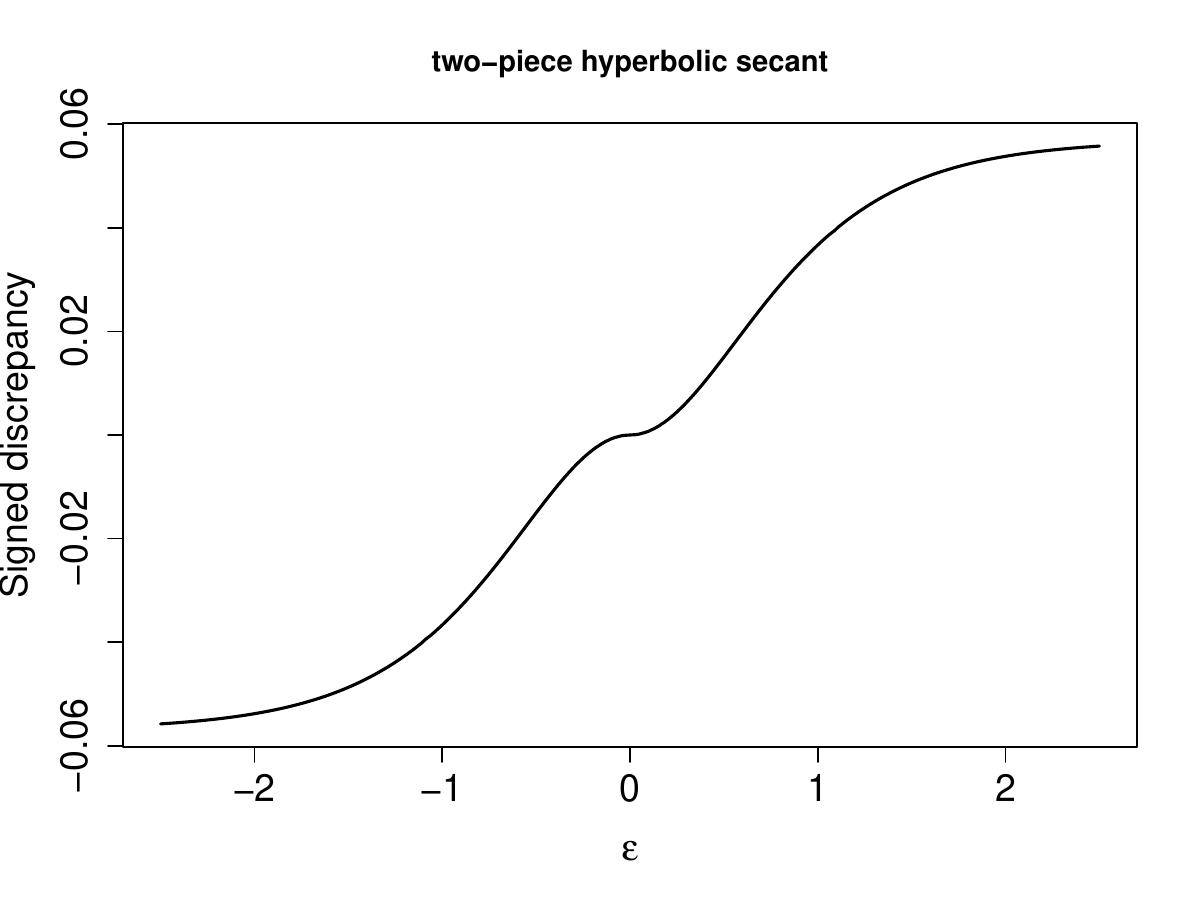} & 
\includegraphics[width=0.3\textwidth]{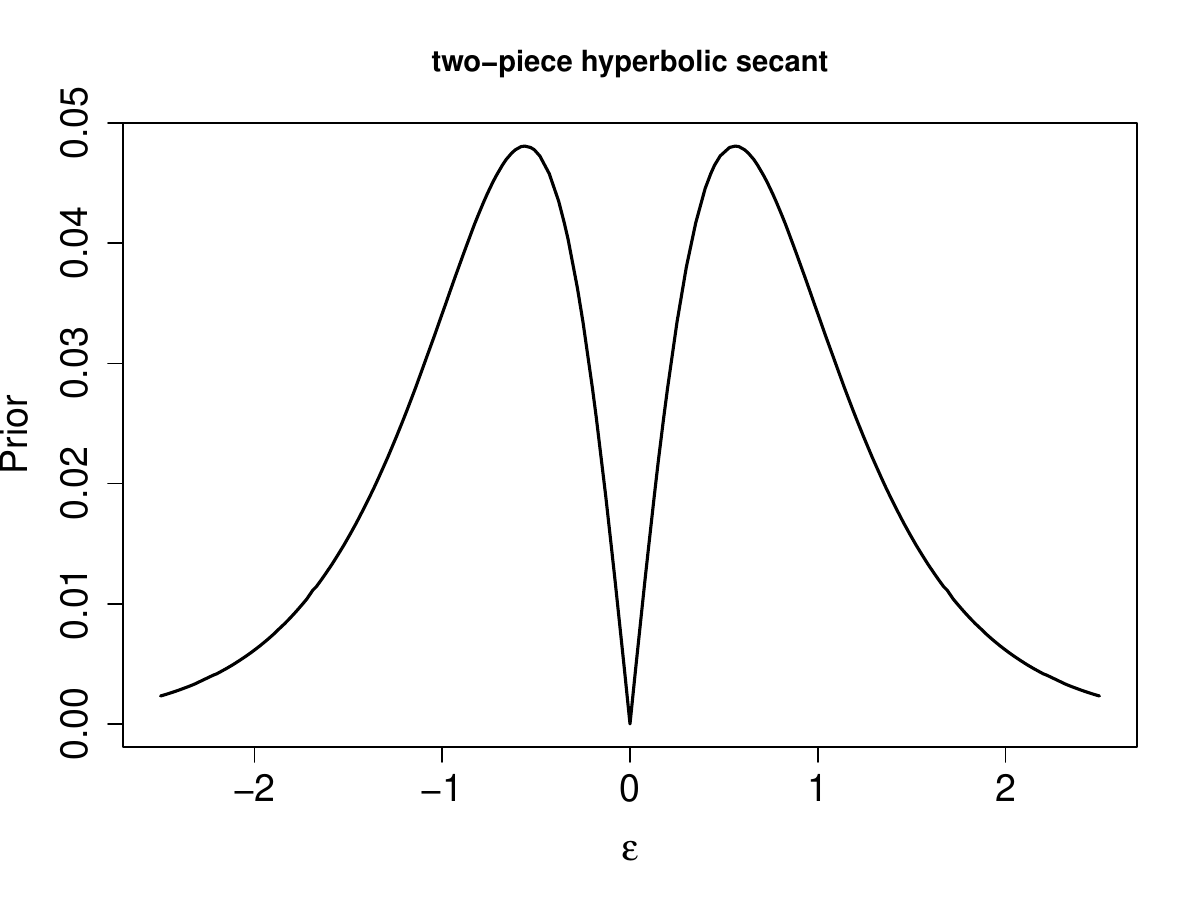} \\
 (a) & (b) & (c)\\
 \end{tabular}
    \caption{Two-piece hyperbolic secant distribution: (a) Discrepancy measure, (b) signed discrepancy measure, and (c) MOOMIN prior.}
    \label{fig:tphs_min}
\end{figure}
} 

\end{document}